\documentclass[twocolumn,english,prl]{revtex4-1}
\usepackage[T1]{fontenc}
\usepackage[latin9]{inputenc}
\usepackage{mathtools}
\usepackage{amsmath}
\usepackage{graphicx}

\makeatletter
\usepackage{braket}
\usepackage{hyperref}

\makeatother

\usepackage{babel}
\begin{document}
\noindent \begin{flushright}
YITP-19-72
\par\end{flushright}

\title{Quantum and Classical Correlations Inside the Entanglement Wedge}

\author{Koji Umemoto$^{a}$}

\affiliation{$^{a}$Center for Gravitational Physics, Yukawa Institute for Theoretical
Physics, Kyoto University, Kitashirakawa Oiwakecho, Sakyo-ku, Kyoto
606-8502, Japan}

\date{\today}
\begin{abstract}
We show that the entanglement wedge cross section (EWCS) can become
 larger than the quantum entanglement measures such as the entanglement of formation in the AdS/CFT correspondence. We then discuss a series of holographic duals to the optimized
correlation measures, finding a novel geometrical measure of correlation, the \textit{entanglement
wedge mutual information} (EWMI), as the dual of the $Q$-correlation. We prove that the EWMI satisfies
the properties of the $Q$-correlation as well as the strong superadditivity, and that it can become larger than the 
entanglement measures. These results imply that both of the EWCS and the EWMI capture more than quantum entanglement in the entanglement wedge, which enlightens a potential
role of classical correlations in holography. 
\end{abstract}
\maketitle

\section{Introduction}

Quantum entanglement has provided a key tool to study various aspects
of modern physics from condensed matter theory to the black hole evaporation.
In the AdS/CFT correspondence \citep{Maldacena99:TheLargeNlimit,GubserKlebanovPolyakov98:GaugeTheoryCorrelators,Witten98:Anti-deSitter},
quantum entanglement also plays a central role in the investigation
of how the bulk geometrical data are encoded in the boundary field
theory \citep{Maldacena03:Eternalblackhole,VanRaamsdonk10:Buildingup,MaldacenaSusskind13:Coolhorizons,Takayanagi18:HolographicSpacetimes,VanRaamsdonk18:Buildingup}.
The Ryu-Takayanagi formula \citep{RyuTakayanagi06:Holographic,LewkowyczMaldacena13:GeneralizedGravitational} (or the Hubeny-Rangamani-Takayanagi formula \citep{HubenyRangamaniTakayanagi07:ACovariantholographic,DongLewkowyczRangamani16:DerivingCovariant} for covariant cases)
tells us that the von Neumann entropy associated with a spacial subregion $A$
in CFTs $S_{A}\equiv S(\rho_{A})=-{\rm Tr}\rho_{A}\log\rho_{A}$
is equivalent to the area of codimension-2 minimal surface $\gamma_{A}$ which is anchored
on the entangling surface $\partial A$ and homologous to
$A$, 
\begin{equation}
S_{A}=\min_{\gamma_{A}}\frac{{\rm Area}(\gamma_{A})}{4G_{N}},\label{eq:RTformula}
\end{equation}
at the leading order of the large $N$ limit. The von Neumann entropy $S_{A}$ is commonly called
the entanglement entropy (EE) because this quantifies an amount of quantum
entanglement between $A$ and its complement $A^{c}$ when the total
state is pure \citep{DonaldHorodeckiRudolph01:TheUniqueness}. For
mixed states, however, the von Neumann entropy no longer deserves
to be a measure of correlation, and thus we need to find another geometrical way to measure correlations.

A generalization of the Ryu-Takayanagi surface, the entanglement
wedge cross section (EWCS), was introduced in \citep{UmemotoTakayanagi18:Entanglementofpurificationthroughholographicduality,NguyenDevakulHalbaschZaletelSwingle17:Enanglementofpurification}
as the minimal cross section of the entanglement wedge \citep{Wall12:Maximinsurfaces,CzechKarczmarekNogueiraVanRaamsdonk12:TheGravityDual,HeadrickHubenyLawrenceRangamani14:Causalityandholographic}.
This is a geometrical measure of correlations between the boundary subsystems connected
by the entanglement wedge which are usually in mixed states. Thus the
EWCS in boundary theories is expected to be dual to some correlation measure which is
a generalization of EE for mixed states.

The EWCS was originally conjectured to be the dual of the entanglement
of purification (EOP) \citep{TerhalHorodeckiLeungDiVincenzo02:TheEntanglementofPurification},
based on agreements of their various information-theoretic properties
\citep{UmemotoTakayanagi18:Entanglementofpurificationthroughholographicduality,NguyenDevakulHalbaschZaletelSwingle17:Enanglementofpurification}
as well as compatibility with the tensor network description of AdS/CFT
\citep{Swingle09:EntanglementRenormalization,MiyajiTakayanagi15:SurfaceStateCorrespondence}.
The proposal has passed further consistency checks in the multipartite
generalization \citep{UmemotoZhou18:Entanglement} and in the conditional
generalization \citep{BaoHalpern17:HolographicInequalities,BaoHalpern18:Conditional}.
Refer to \citep{HiraiTamaokaYokoya18:TowardsEntanglementofPurification,EspindolaGuijosaPedraza18:EntanglementwedgereconstructionandEntanglementofPurification,BaoChatwin-DaviesRemmem18:EntanglementofPurificationandMultiboundaryWormhole,YangZhangLi18:HolographicEntanglementofPurificationforthermofielddouble,CaputaMiyajiTakayanagiUmemoto18:HolographicEntanglementofPurificationFromConformalFieldTheories,LiuLingNiuWu19:EntanglementofPurificationinHolographicSystems,Guo19:EntanglementofPurificationandProjectiveMeasurement,BhattacharyyaJahnTakayanagiUmemoto19:EntanglementofPurificationinManybody,GhodratiKuangWangZhangZhou19:TheconnectionbetweenholographicentanglementandComplexityofPurification,Prudenziati19:AgeodesicWittendiagramdescriptionofholographicentanglement,BaoPeningtonSorceWall19:HolographidTensorNetworksinFullAdSCFT,BabaeiVelniMohammadiMozaffarVahidinia19:SomeAspectsofEntanglementwedgecrosssection,Ota19:CommentsonholographicentanglementsincutoffAdS,JokelaPonni19:Notesonentanglementwedgecrosssections,Guo19:EntanglementofPurificationanddisentanglement,KusukiTamaoka19:DynamicsofEntanglementWedge}
for recent progress.

Surprisingly, several correlation measures other than EOP have been
shown to be essentially equal to the EWCS with appropriate coefficients, including the logarithmic negativity \citep{Kudler-FlamRyu18:EntanglementNegativityandMinimalEntanglementWedge,Kudler-FlamMacCormackRyu19:HolographicentanglementcontourBitthreadsandtheEntanglementTsunami,Kudler-FlamNozakiRyuTan19:QuantumvsClassical,KusukiKudler-FalmRyu19:DerivationofHolographicNegativity},
the odd entropy \citep{Tamaoka18:EntanglementWedgeCrossSection},
and the reflected entropy \citep{DuttaFaulkner19:Acanonicalpurification}.
With the monogamy of holographic mutual information
\citep{HaydenHeadrickMaloney11:HolographicMutualInformation} in mind, which strongly suggests
that quantum entanglement dominates holographic correlations, we may speculate that some axiomatic measure of quantum entanglement
(see e.g. \citep{HHHH:QuantumEntanglement}) would also be equivalent to
the EWCS in holographic CFTs.

In this paper, however, we present a no-go theorem in this direction:
the EWCS is \textit{not} dual of various entanglement measures. Furthermore, we show that the EWCS can be strictly larger than various
entanglement measures at the leading order $O(N^{2})$. It is particularly shown in a holographic configuration near to the saturation of the Araki-Lieb inequality \citep{HubenyMaxfieldRangamaniTonni13:Holographicentanglementplateaux,Headrick14:Generalpropertiesof}.
We also point out that the EWCS is also larger than another type of
quantum correlation, the quantum discord \citep{HendersonVedral01:ClassicalQuantumandTotal,OllivierZurek01:QuantumDiscord}.
It implies that the EWCS captures more than quantum entanglement in the entanglement wedge, and it must be sensitive to classical correlations as well. 

Next, we introduce a series of holographic duals for the optimized
correlation measures, which are akin to the EOP. This class includes two entanglement
measures; the squashed entanglement \citep{ChristandlWinter04:SquashedEntanglement}
and the conditional entanglement of mutual information (CEMI) \citep{YangHorodeckiWang08:AnAdditive},
and three total correlation measures; the EOP, the $Q$-correlation,
and the $R$-correlation \citep{LevinSmith19:OptimizedMeasures}.
We show that the CEMI reduces to half of the holographic mutual information and the $R$-correlation does to the EWCS, when they are optimized over the geometrical extensions. These two duals thus do not lead to new geometrical object in the bulk.

However, we find that
the holographic dual of the $Q$-correlation provides us with a new
bulk measure of correlation inside the entanglement wedge, which we call the \textit{entanglement
wedge mutual information} (EWMI). This quantity appropriately satisfies all
of the properties of the $Q$-correlation, as well as the strong superadditivity like the EWCS. Furthermore, we show
that the EWMI can also strictly become larger
than the various quantum correlation measures in the same holographic configurations. It again implies that classical
correlations are included in holographic correlations and they are geometrically encoded in the
entanglement wedge.

This paper is organized as follows: In section 2, we review the basic
notion of the EWCS and information-theoretic correlation measures.
In section 3, we show that the EWCS is strictly larger
than various measures of quantum correlation in a holographic configuration near to the saturation of 
the Araki-Lieb inequality. In section 4, we argue
 holographic duals of the optimized correlation measures, introduce the EWMI, 
and discuss the aspects of the EWMI.
In section 5, we discuss some future problems. In the appendix, we prove new inequalities of the multipartite
EOP and the multipartite EWCS, complementing the work
of \citep{UmemotoZhou18:Entanglement}. 

\textit{Note:} We became aware that an independent work \citep{LevinSmith19:Toappear}
which partially overlaps with the present paper will appear soon.

\section{Preliminaries}

\subsection{Entanglement Wedge Cross Section}

\begin{figure}
\includegraphics[scale=0.14]{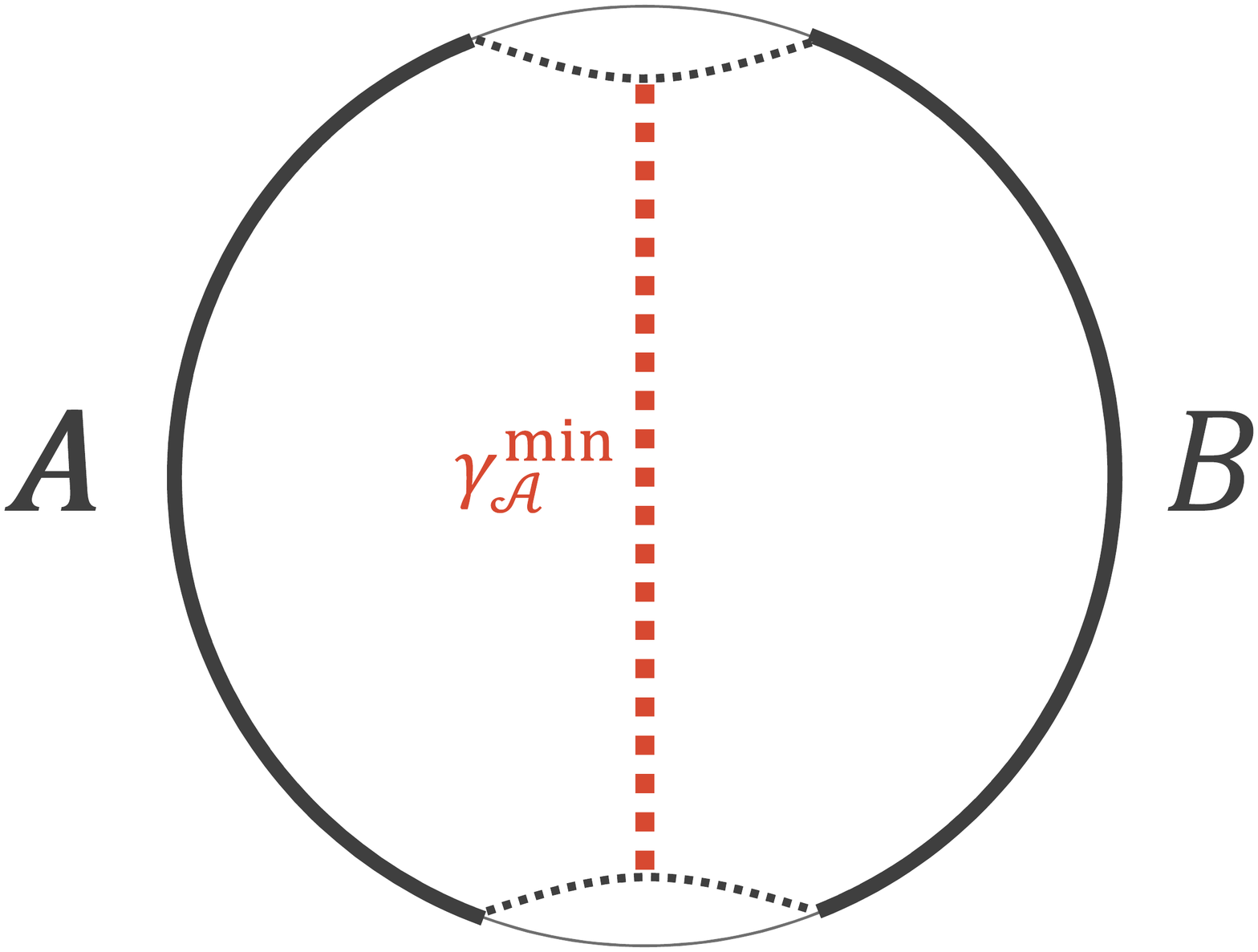}
\includegraphics[scale=0.14]{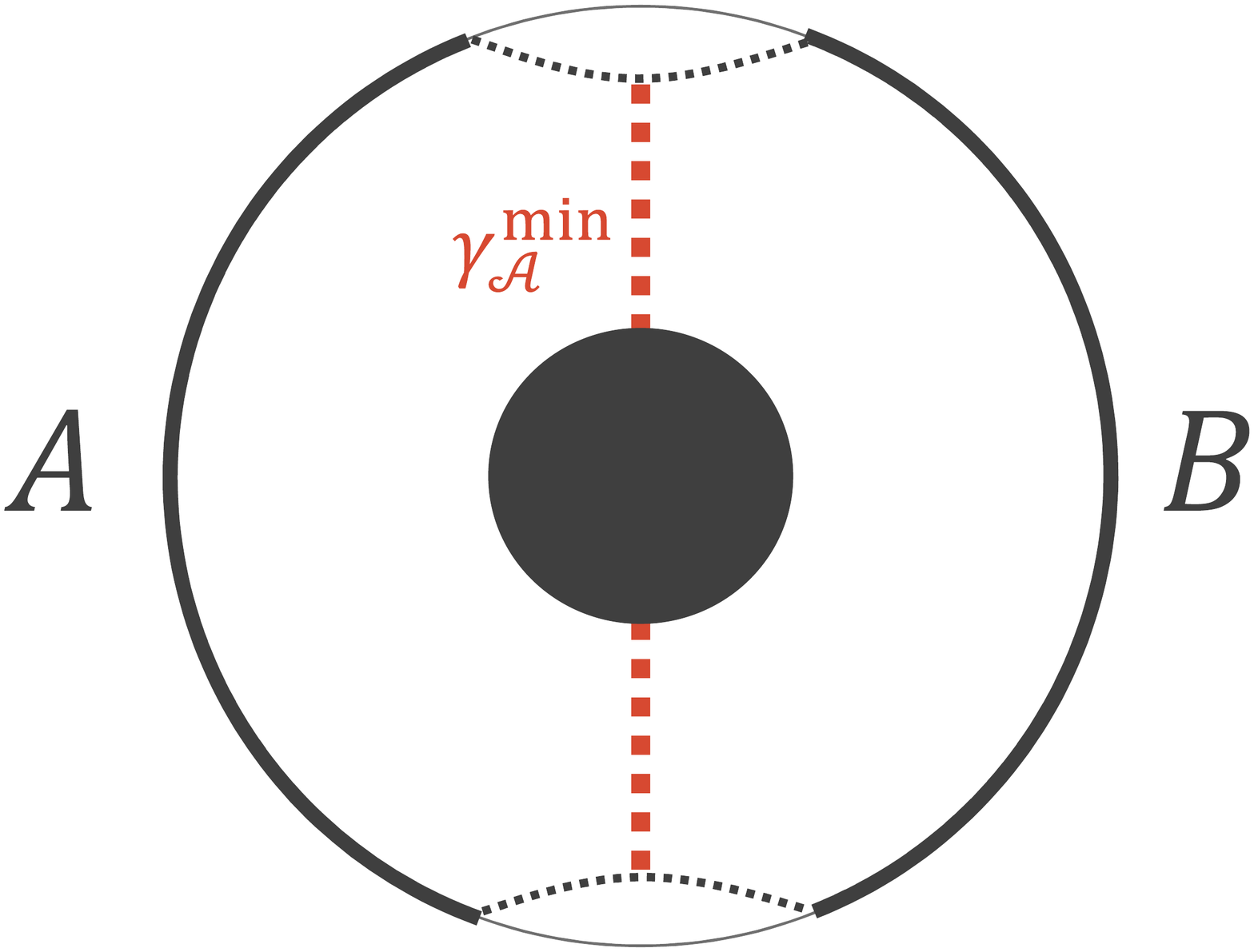}
\caption{\label{fig:EWCS} The EWCS (red dashed lines) on a time slice of the entanglement wedge. }
\end{figure}

In the present paper we deal with static spacetime for simplicity
(a generalization to non-static spacetime is straightforward using
the HRT-formula \citep{HubenyRangamaniTakayanagi07:ACovariantholographic,DongLewkowyczRangamani16:DerivingCovariant}
instead of the RT-formula). The boundary subsystems are denoted by
$A$ and $B$ and the entanglement wedge of $AB\equiv A\cup B$ (on
a canonical time slice) is denoted by $\mathcal{M}_{AB}$ \citep{Wall12:Maximinsurfaces,CzechKarczmarekNogueiraVanRaamsdonk12:TheGravityDual,HeadrickHubenyLawrenceRangamani14:Causalityandholographic}.
Given an entanglement wedge $\mathcal{M}_{AB}$, we may define the
minimal cross section as follows \citep{UmemotoTakayanagi18:Entanglementofpurificationthroughholographicduality,NguyenDevakulHalbaschZaletelSwingle17:Enanglementofpurification}: 

Suppose the boundary of $\mathcal{M}_{AB}$ is divided into two ``subsystems''
$\mathcal{A}$ and $\mathcal{B}$ i.e. $\partial\mathcal{M}_{AB}=\mathcal{A}\cup\mathcal{B}$
under the condition $\mathcal{A}=A\cup A'$, $\mathcal{B}=B\cup B'$.
We include the asymptotic AdS boundary and (if it exists) black hole
horizon in the boundary of $\mathcal{M}_{AB}$. The EWCS of $\mathcal{M}_{AB}$,
$E_{W}(A:B)$, is defined as the minimum of the holographic entanglement
entropy $S_{\mathcal{A}}$ optimized over all possible partitions (Fig.\ref{fig:EWCS})
\begin{align}
E_{W}(A:B):= & \min_{\mathcal{A}:\mathcal{\partial M}_{AB}=\mathcal{A}\cup\mathcal{B},\ A\subset\mathcal{A},\ B\subset\mathcal{B}}S_{\mathcal{A}}\\
 & =\min_{\gamma_{\mathcal{A}}}\frac{{\rm Area}(\gamma_{\mathcal{A}})}{4G_{N}},\label{eq:EWCS}
\end{align}
where $\gamma_{\mathcal{A}}$ is the RT-surface of $\mathcal{A}$.
It gives a generalization of (\ref{eq:RTformula}) for mixed states
in the sense that $\gamma_{\mathcal{A}}$ reduces to the usual RT-surface
when $\rho_{AB}$ is a pure state. The EWCS always satisfies the inequalities
$\frac{1}{2}I(A:B)\leq E_{W}(A:B)\leq\min\{S_{A},S_{B}\}$, where
$I(A:B)\coloneqq S_{A}+S_{B}-S_{AB}$ is the mutual information. The
above definition can be generalized to $n$-partite subsystems \citep{UmemotoZhou18:Entanglement}.
Remarkably, the EWCS can be regarded as a generalization of the area
of a wormhole horizon in the canonical purification \citep{DuttaFaulkner19:Acanonicalpurification}. 

\subsection{Information-theoretic Correlation Measures}

The EWCS was originally conjectured to be dual to the entanglement
of purification (EOP) at the leading order $O(N^{2})$. The EOP is
defined for a bipartite state $\rho_{AB}$ by \citep{TerhalHorodeckiLeungDiVincenzo02:TheEntanglementofPurification}
\begin{equation}
E_{P}(A:B):=\min_{\ket{\psi}_{AA'BB'}}S_{AA'}=\frac{1}{2}\min_{\ket{\psi}_{AA'BB'}}I(AA':BB'),\label{eq:EOP}
\end{equation}
where the minimization is performed over all possible purifications.
The information-theoretic properties of EOP \citep{TerhalHorodeckiLeungDiVincenzo02:TheEntanglementofPurification,BagchiPati15:MonogamyPolygamy}
are proven for the EWCS geometrically, including the multipartite
cases \citep{UmemotoZhou18:Entanglement}. Moreover, the surface/state
correspondence of the tensor network description \citep{MiyajiTakayanagi15:SurfaceStateCorrespondence}
allows us to find a heuristic derivation of $E_{W}=E_{P}$ \citep{UmemotoTakayanagi18:Entanglementofpurificationthroughholographicduality}. 

The EOP, the mutual information, the $Q$-correlation and $R$-correlation
\citep{LevinSmith19:OptimizedMeasures} (which will be defined in
the section \ref{sec:HDofOpt}) are monotonically non-increasing under
local operations (LO), but may increase by classical communication
(CC). We call such non-negative quantities on $\rho_{AB}$ as (bipartite)
\textit{total correlation measures}. On the other hand, entanglement
measures are defined by monotonicity under LOCC. There is a class
of entanglement measures which satisfy additional axioms such as asymptotic
continuity, which we call (bipartite) \textit{axiomatic entanglement
measures }(see e.g. \citep{HHHH:QuantumEntanglement}). There are
various choices of additional axioms one can impose. In what follows
we make a somewhat minimal requirement motivated by the uniqueness
theorem \citep{DonaldHorodeckiRudolph01:TheUniqueness}: They coincide
with EE for pure states. This may be regarded as a normalization condition
for different measures. Such a class includes, for instance, the distillable
entanglement $E_{D}$ \citep{BennettDiVincenzoSmolinWootters96:MixedstateEntanglement,Rains99:Rigoroustreatmentofdistillable},
the squashed entanglement $E_{sq}$ \citep{Tucci02:Entanglement,ChristandlWinter04:SquashedEntanglement},
the conditional entanglement of mutual information $E_{I}$ \citep{YangHorodeckiWang08:AnAdditive},
the relative entropy of entanglement $E_{RE}$ \citep{VedralPlenioRippinKnikght97:QuantifyingEntanglement},
the entanglement cost $E_{C}$ \citep{BennettDiVincenzoSmolinWootters96:MixedstateEntanglement,HaydenHorodeckiTerhal01:TheAsymptoticEntanglementCost},
and the entanglement of formation $E_{F}$ \citep{BennettDiVincenzoSmolinWootters96:MixedstateEntanglement}.
There is another measure of quantum correlation, called the quantum
discord $D$ \citep{HendersonVedral01:ClassicalQuantumandTotal,OllivierZurek01:QuantumDiscord}.
It captures wider types of quantum correlation than quantum entanglement,
and coincides with EE for pure states. 

\section{EWCS is Not Dual of Axiomatic Entanglement Measures}

First of all, we can use the generic upper bounds $E_{D},E_{sq},E_{I},E_{RE},D\leq I$
to exclude $E_{D},\ E_{sq},\ E_{I},\ E_{RE}$ and $D$ as a dual candidate
of $E_{W}$, since $E_{W}(A:B)>I(A:B)$ can be observed near to the
$O(1)$ phase transition of $I(A:B)$ \citep{UmemotoTakayanagi18:Entanglementofpurificationthroughholographicduality}.
It already gives us intuition that the entanglement (or quantum correlation)
measures are usually less than the EWCS in holographic CFTs. In this
way, however, we cannot exclude $E_{C}$ and $E_{F}$ since they may
exceed $I(A:B)$ (they can be greater than $I(A:B)/2$ \cite{LiLuo07:Totalversusquantum}). In order to do that, we consider another particular holographic
setup as follows. 

\subsection{The EWCS in the Araki-Lieb Transition\label{subsec:EWCS_ALtransition}}

\begin{figure}
\includegraphics[viewport=0bp 150bp 842bp 595bp,scale=0.25]{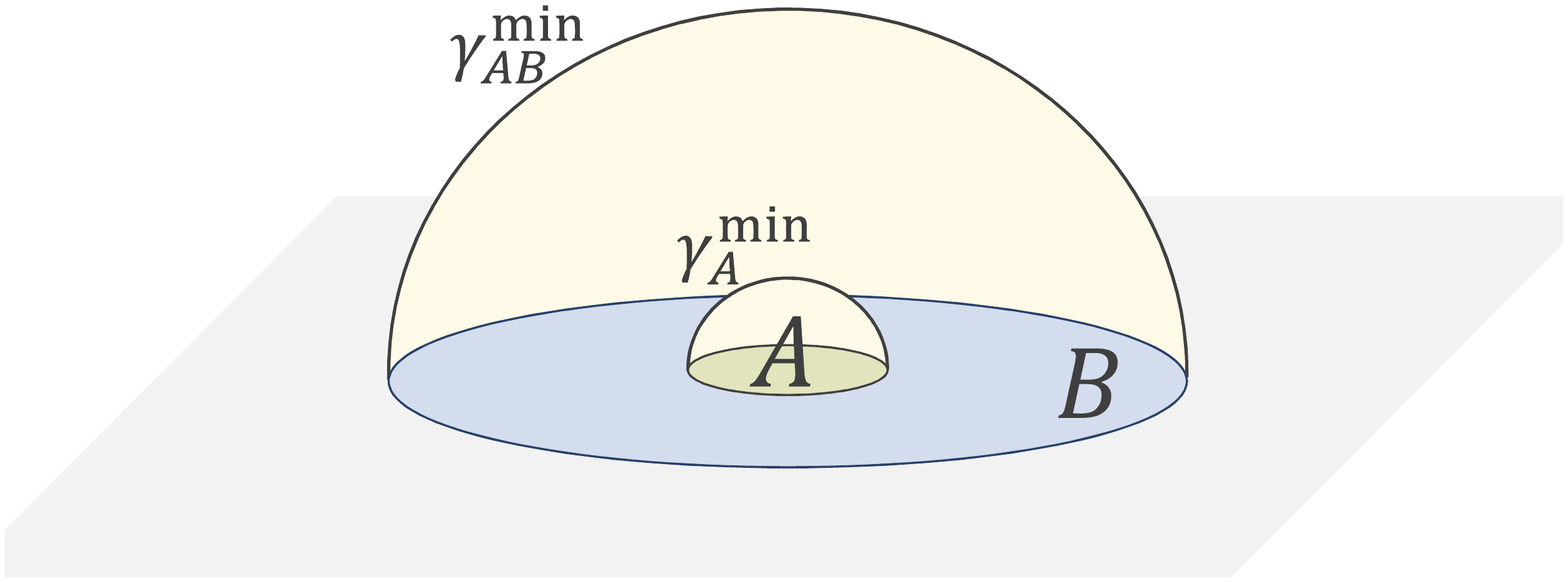}
\caption{\label{fig:ALsat_config} A holographic configuration for which the
Araki-Lieb inequality is saturated $S_{A}+S_{AB}=S_{B}$. }
\end{figure}

One of outstanding characteristics of holographic CFTs is the fact
that the Araki-Lieb inequality,
\begin{equation}
S_{A}+S_{AB}\geq S_{B},\label{eq:ArakiLiebineq}
\end{equation}
can be saturated at the leading order $O(N^{2})$ in some particular
configurations \citep{HubenyMaxfieldRangamaniTonni13:Holographicentanglementplateaux,Headrick14:Generalpropertiesof}.
It is typically realized by a subsystem $A$ completely surrounded
by sufficiently large $B$ (Fig.\ref{fig:ALsat_config})
. Though the following discussion is valid for the more generic setups,
we focus on a configuration in Poincar\'{e} ${\rm AdS}_{3}$ with
the metric 
\begin{equation}
ds^{2}=\frac{dz^{2}-dt^{2}+dx^{2}}{z^{2}}.
\end{equation}

Suppose the subsystems $A$ and $B$ are given by $A=[-a,a],\ B=[-b,-a]\cup[a,b]\equiv B_{1}\cup B_{2}$
for $0<a<b$ w.l.o.g. We also define the relative size of subsystems
by $p\equiv\frac{a}{b}$ for $p\in(0,1)$. The mutual information
$I(A:B)$ exhibits a phase transition due to that of $S_{B}=S_{B_{1}B_{2}}$
depending on the relative size $p$. The connected phase $I(B_{1}:B_{2})>0$
is preferred if $p$ is small, and the disconnected phase $I(B_{1}:B_{2})=0$
is if it is large. Thus $I(A:B)$ can be computed as 
\begin{align}
I(A:B) & =S_{A}+S_{B}-S_{AB}\nonumber \\
 & =\min\{\frac{2c}{3}\log\frac{2a}{\epsilon},\frac{2c}{3}\log\frac{\sqrt{a/b}(b-a)}{\epsilon}\},
\end{align}
where $c$ is the central charge of holographic $2$d CFTs and $\epsilon$
is the UV cutoff. It is divergent since we are taking the adjacent
limit. The phase transition point of $I(A:B)$ can be read off as
\begin{equation}
p_{{\rm MI}}^{*}\equiv\frac{a_{{\rm MI}}^{*}(b)}{b}=3-2\sqrt{2}.
\end{equation}
The Araki-Lieb inequality is saturated for $0<p<p_{{\rm MI}}^{*}$
but not for $p_{{\rm MI}}^{*}<p<1$. 

\begin{figure}
\includegraphics[scale=0.2]{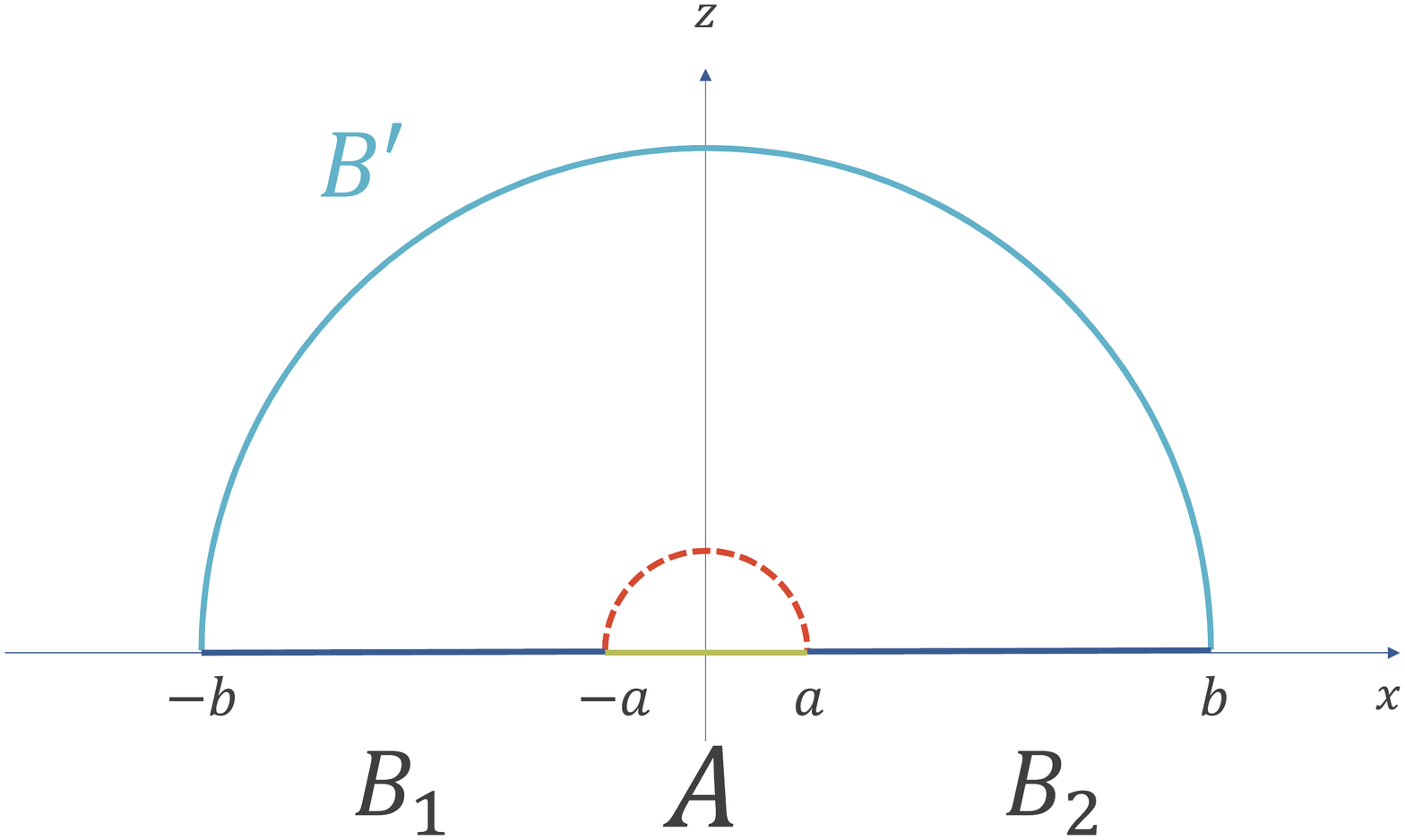}
\includegraphics[scale=0.2]{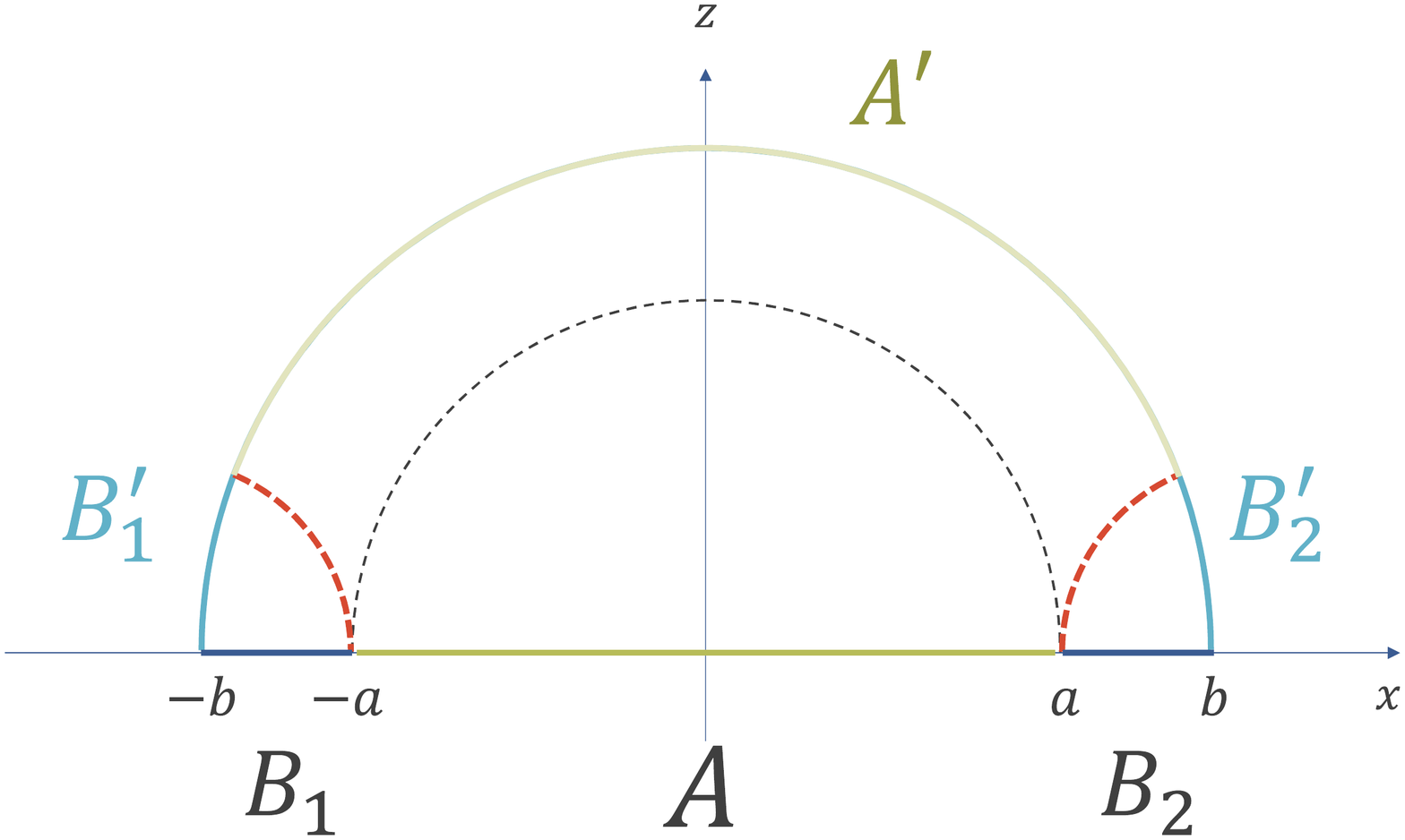}
\caption{\label{fig:EWCSALtransition} The two configurations of the EWCS $E_{W}(A:B)=E_{W}(A:B_{1}B_{2})$,
denoted by the orange dashed line, for the symmetric setup in the Poincar\'{e}
${\rm AdS}_{3}$. The left (right) configuration is preferred when
the relative size $p<p_{{\rm EW}}^{*}$ ($p>p_{{\rm EW}}^{*}$). The
primed symbols allocated on the upper semi-circle denote the partition
in (\ref{eq:EWCS}) with $\mathcal{A}=A\cup A'$ and $\mathcal{B}=B_{1}\cup B_{2}\cup B_{1}'\cup B_{2}'$.}
\end{figure}

The EWCS also exhibits a phase transition depending on the relative
size of $A$ (Fig.\ref{fig:EWCSALtransition}). The formula for the
EWCS in Poincar\'{e} ${\rm AdS}_{3}$ is given in \citep{UmemotoTakayanagi18:Entanglementofpurificationthroughholographicduality}
by
\begin{align}
E_{W} & =\min\{S_{A},\ 2E_{W}(AB_{1}:B_{2})\}\nonumber \\
 & =\min\{\frac{c}{3}\log\frac{2a}{\epsilon},\ \frac{c}{3}\log[\frac{b^{2}-a^{2}}{b\epsilon}]\}.
\end{align}
The phase transition of EWCS therefore happens at 
\begin{equation}
p_{{\rm EW}}^{*}\equiv\frac{a_{{\rm EW}}^{*}(b)}{b}=\sqrt{2}-1.
\end{equation}

\begin{figure}
\includegraphics[scale=0.3]{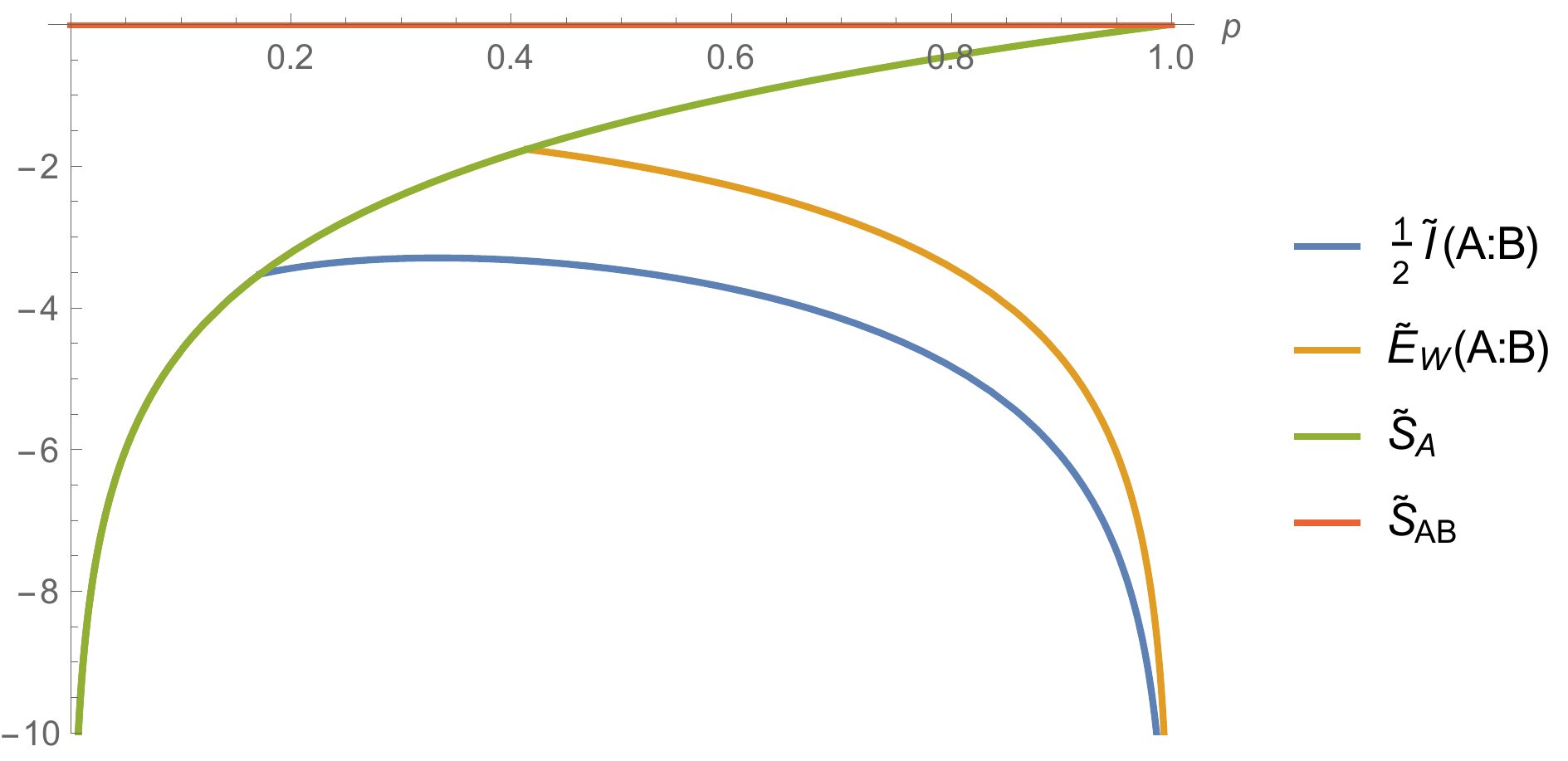}
\caption{\label{fig:MIandEWCSforALtransition} Half of the mutual information
and the EWCS for the Araki-Lieb transition (normalized by subtracting
$S_{AB}$).}
\end{figure}

The phase transition points of the mutual information and the EWCS
do not match, and the strict inequality $p_{{\rm MI}}^{*}<p_{{\rm EW}}^{*}$
holds (Fig.\ref{fig:MIandEWCSforALtransition}).
 This means that the EWCS saturates its upper bound $E_{W}(A:B)=S_{A}$
while the Araki-Lieb inequality is \textit{not} saturated for $p\in(p_{{\rm MI}}^{*},p_{{\rm EW}}^{*})$.
This observation provides us a crucial benchmark: A correlation measure
$E$ \textit{can not} be dual to the EWCS if $E(A:B)=S_{A}$ automatically
implies the saturation of the Araki-Lieb inequality $S_{A}+S_{AB}=S_{B}$.

The Araki-Lieb inequality is also holographically saturated in the
global BTZ black hole
\begin{align}
ds^{2} & =\frac{f^{-1}(z)dz^{2}-f(z)dt^{2}+dx^{2}}{z^{2}},\\
f(z) & =1-\frac{z^{2}}{z_{H}^{2}},
\end{align}
with the inverse temperature $\beta=2\pi z_{H}$ and the periodic
boundary condition $x\simeq x+2\pi$. We choose $A=[-l/2,l/2]$ for
$l\in(0,\pi)$ and $B$ as the reminder. It exhibits the Araki-Lieb
saturation when the size of $A$ is small enough. We find the phase
transition points (see e.g. \citep{HubenyMaxfieldRangamaniTonni13:Holographicentanglementplateaux,UmemotoTakayanagi18:Entanglementofpurificationthroughholographicduality,NguyenDevakulHalbaschZaletelSwingle17:Enanglementofpurification})
\begin{align}
l_{{\rm MI}}^{*}(z_{H}) & =\pi-z_{H}\log\cosh(\frac{\pi}{z_{H}}),\\
l_{{\rm EW}}^{*}(z_{H}) & =2z_{H}\log(1+\sqrt{2}).
\end{align}
This leads to $l_{{\rm MI}}^{*}(z_{H})<I_{{\rm EW}}^{*}(z_{H})$ for
any $z_{H}>0$, which confirms the above conclusion. 

\subsection{The Axiomatic Entanglement Measures and the Araki-Lieb Saturation}

We can now cite the following fact: the entanglement of formation
saturates its upper bound $E_{F}(A:B)\leq S_{A}$ if and only if the
Araki-Lieb inequality is saturated \citep{XiLuWangLi12:NecessaryandSufficient}.
This immediately means that $E_{F}$ is not dual of $E_{W}$ from
the above observation. Furthermore, this statement can be generalized
to other measure $E$ which satisfies (i) the monotonicity $E(A:B_{1}B_{2})\geq E(A:B_{1})$,
(ii) $E(A:B)=S_{A}=S_{B}$ for pure states, and (iii) $E\leq E_{F}$.
Indeed, the saturation $E(A:B)=S_{A}$ leads to $E_{F}(A:B)=S_{A}$
from (iii), which is equivalent to the Araki-Lieb saturation. The
opposite is shown by using the unique structure (up to isometries) of states which
saturate the Araki-Lieb inequality \citep{ZhangWu11:Onconjectures}
\begin{equation}
\rho_{AB}=\ket{\psi}\bra{\psi}_{AB_{L}}\otimes\rho_{B_{R}},\label{eq:ALsat_state}
\end{equation}
where the Hilbert space of $B$ is decomposed into $\mathcal{H}_{B}=\mathcal{H}_{B_{L}}\otimes\mathcal{H}_{B_{R}}$.
This form with (i) and (ii) leads to the saturation by $S_{A}\geq E(A:B_{L}B_{R})\geq E(A:B_{L})=S_{A}$.
This class of correlation measures especially includes the entanglement
cost $E_{C}$, and the other entanglement measures mentioned at the
beginning of this section as well. Thus we can also exclude $E_{C}$
and the other measures as a dual candidate of $E_{W}$.

In addition, it also means that this class of entanglement measures
must be \textit{strictly less} than $E_{W}$ for the states of $p\in(p_{{\rm MI}}^{*},p_{{\rm EW}}^{*})$.
Since there seems to be no reason to believe that these states are
singular among the holographic states, we argue that the EWCS is generically
larger than these entanglement measures in holographic CFTs (unless
the Araki-Lieb inequality is saturated). 

\subsection{Interpretation from the Holographic Entanglement of Purification}

By contrast, the EOP evades the above criteria since (iii) does not
hold, and still deserves consideration as a possible dual of the EWCS.
In addition, there exists a class of states for which $E_{P}(A:B)=S_{A}$,
but the Araki-Lieb inequality is not saturated \citep{CheistandlWinter05:Uncertainty}.
Similarly, the logarithmic negativity, the odd entropy, and the reflected
entropy do not satisfy (iii), and we expect that they also coincide
with $S_{A}$ with an appropriate coefficient for some mixed states
without the Araki-Lieb saturation. 

Furthermore, we can understand the behavior of the EWCS through the
Araki-Lieb transition based on the surface/state correspondence. First,
we note a remarkable equality which holds after the phase transition
$p>p_{{\rm EW}}^{*}$, 
\begin{equation}
E_{W}(A:B_{1}B_{2})=E_{W}(AB_{1}:B_{2})+E_{W}(AB_{2}:B_{1}).\label{eq:TriEqEWCS}
\end{equation}
This equation can be explained as follows: the two configurations
$p>p_{{\rm EW}}^{*}$ and $p<p_{{\rm EW}}^{*}$ are equivalent to
whether the correlation\textbf{ }$I(\mathcal{B}_{1}:\mathcal{B}_{2})$
vanishes or not. For $p>p_{{\rm EW}}^{*}$, we see $I(\mathcal{B}_{1}:\mathcal{B}_{2})=0$,
and it immediately leads to the unique form (up to isometries on $\mathcal{H}_{\mathcal{A}}$)
of any purification \citep{Uhlmann76:Thetransitionprobabilityeltit}
\begin{equation}
\ket{\psi}_{\mathcal{A}\mathcal{B}_{1}\mathcal{B}_{2}}=\ket{\phi^{1}}_{\mathcal{A}\mathcal{B}_{1}}\otimes\ket{\phi^{2}}_{\mathcal{A}\mathcal{B}_{2}}.\label{eq:TriEqPurification}
\end{equation}
This form of optimal purification, common to each of the three EWCSs,
clearly establishes the equality (\ref{eq:TriEqEWCS}). 

On the other hand, if $p<p_{{\rm EW}}^{*}$, remaining correlation
$I(\mathcal{B}_{1}:\mathcal{B}_{2})>0$ will drastically change the
structure of purifications from (\ref{eq:TriEqPurification}). In
this case, the optimal purification would be simply given by the standard
purification \citep{TerhalHorodeckiLeungDiVincenzo02:TheEntanglementofPurification}
i.e. setting $A'$ as empty. In this sense, the phase transition point
$p_{{\rm EW}}^{*}$ is thus understood as a point at which the standard
purification switches with the decoupled purification (\ref{eq:TriEqPurification})
as the optimal purification.

\section{Holographic Duals of the Optimized Correlation Measures\label{sec:HDofOpt}}

We observed that the EWCS can not be the dual of any axiomatic entanglement
measures. Then a natural question is as follows: Is there any axiomatic entanglement
measure which deserves a geometrical dual?

\subsection{Holographic Dual of the Optimized Entanglement Measures\label{subsec:HDofEnt}}

Here we discuss two possible candidates: the squashed entanglement
$E_{sq}$ \citep{ChristandlWinter04:SquashedEntanglement} and the
conditional entanglement of mutual information $E_{I}$ \citep{YangHorodeckiWang08:AnAdditive}.
Their definitions are reminiscent of the EOP (\ref{eq:EOP}). The
squashed entanglement is defined as 
\begin{align}
E_{sq}(A:B) & :=\frac{1}{2}\min_{\rho_{ABE}}I(A:B|E)\\
 & =\frac{1}{2}I(A:B)-\frac{1}{2}\max_{\rho_{ABE}}I_{3}(A,B,E),
\end{align}
where $\rho_{ABE}$ is an extension such that ${\rm Tr}_{E}\rho_{ABE}=\rho_{AB}$,
and $I_{3}(A,B,C)=S_{A}+S_{B}+S_{C}-S_{AB}-S_{BC}-S_{CA}+S_{ABC}$
is the tripartite information.

We now impose a crucial assumption to find a possible geometrical
dual of $E_{sq}$: Performing the minimization over a class of extensions
which have classical geometrical duals, is sufficient to achieve the
minimum. It implies that the monogamy of mutual information $I_{3}(A,B,E)\leq0$
\citep{HaydenHeadrickMaloney11:HolographicMutualInformation} must
hold for the extensions $\rho_{ABE}$. A holographic dual of the squashed
entanglement is then given by half of the holographic mutual information
\citep{HaydenHeadrickMaloney11:HolographicMutualInformation}, 
\begin{equation}
E_{sq}(A:B)=\frac{1}{2}I(A:B) \label{EsqMI}.
\end{equation}
This is achieved by a trivial extension $E=\emptyset$. This relation
implies that the holographic mutual information should satisfy the
properties of the squashed entanglement, such as the monogamy relation $E_{sq}(A:BC)\geq E_{sq}(A:B)+E_{sq}(A:C)$
\citep{KoashiWinter04:Monogamy} which is generically considered as a characteristic of quantum entanglement. The holographic mutual information
indeed satisfies the monogamy relation as mentioned above. It is worth
noting that the saturation of $E_{sq}\leq\frac{1}{2}I$ occurs if
$\rho_{AB}$ saturates Araki-Lieb inequality, but this is not the
only possibility \citep{CheistandlWinter05:Uncertainty}.

The relation (\ref{EsqMI}), or the monogamy property of the mutual information, 
suggests a striking conclusion: \textit{the mutual information
captures only quantum entanglement in holography, even though it is usually a total
correlation measure} \citep{HaydenHeadrickMaloney11:HolographicMutualInformation}. 

We give support for this argument by elaborating on the conditional
entanglement of mutual information $E_{I}$ \citep{YangHorodeckiWang08:AnAdditive}.
It is defined by
\begin{align}
 & E_{I}(A:B)\nonumber \\
 & :=\frac{1}{2}\min_{\rho_{ABA'B'}}(I(AA':BB')-I(A':B'))\label{eq:CEMI}\\
 & =\frac{1}{2}I(A:B)\nonumber \\
 & \ +\frac{1}{2}\min_{\rho_{ABA'B'}}(I(AA':BB')-I(A:B)-I(A':B')),
\end{align}
where $\rho_{ABA'B'}$ is again any extension of $\rho_{AB}$. It
is an additive measure of quantum entanglement \citep{YangHorodeckiWang08:AnAdditive}.
Suppose the monogamy of mutual information for some geometric extensions
$\rho_{AA'BB'}$ is enough to find the minimum. Then we find $I(AA':BB')-I(A:B)-I(A':B')\geq I(A:B')+I(B:A')\geq0$,
which leads to the holographic dual of the CEMI as half of the holographic
mutual information (with a trivial extension $A'B'=\emptyset$)
\begin{equation}
E_{I}(A:B)=\frac{1}{2}I(A:B).
\end{equation}
It again implies that the holographic mutual information only captures
quantum entanglement. This is in contrast to the EWCS which still
captures classical correlations in holography. Indeed, it was pointed
out in \citep{BhattacharyyaJahnTakayanagiUmemoto19:EntanglementofPurificationinManybody}
that the EOP could be more sensitive to classical correlations than
the mutual information. 

These proposals about $E_{sq}$ and $E_{I}$ are obviously consistent
with the Araki-Lieb transition discussed above, since the holographic
dual of $E_{sq}$ and $E_{I}$ would be the holographic mutual information
itself.

We emphasize the fact that two differently defined measures of entanglement
reduce to the same quantity $\frac{1}{2}I$ in holography. To our
knowledge, there seems to be no obstruction to speculate that the
other entanglement measures such as $E_{C}$ and $E_{F}$ also coincide
with $\frac{1}{2}I$. We leave investigating their holographic duals
as an interesting future work.

\subsection{Holographic Duals of the Optimized Total Correlation Measures\label{subsec:HDofTotal}}

All of the correlation measures $E_{P},\ E_{sq},\ E_{I}$ are defined
as the minimum of a linear combination of von Neumann entropies over
all possible purifications or extensions. This class of correlation
measures is called the \textit{optimized correlation measure}s \citep{LevinSmith19:OptimizedMeasures}.
There are two other such measures, the $Q$-correlation and the $R$-correlation,
introduced in \citep{LevinSmith19:OptimizedMeasures} 
\begin{align}
E_{Q}(A:B) & \coloneqq\frac{1}{2}\min_{\rho_{ABE}}(S_{A}+S_{B}+S_{AE}-S_{BE})\\
 & \equiv\min_{\rho_{ABE}}f^{Q}(A,B,E).\\
E_{R}(A:B) & \coloneqq\frac{1}{2}\min_{\rho_{ABE}}(S_{AB}+2S_{AE}-S_{ABE}-S_{E})\\
 & \equiv\min_{\rho_{ABE}}f^{R}(A,B,E).
\end{align}
The symmetry between $A$ and $B$ becomes obvious in the equivalent
expression in terms of purifications ($E\equiv A'$), 
\begin{align}
 & E_{Q}(A:B)\nonumber \\
 & =\frac{1}{2}\min_{\ket{\psi}_{AA'BB'}}(S_{A}+S_{B}+\frac{S_{AA'}+S_{BB'}-S_{BA'}-S_{AB'}}{2})\\
 & \equiv\min_{\ket{\psi}_{AA'BB'}}f^{Q}(A,A',B,B').\\
 & E_{R}(A:B)\nonumber \\
 & =\frac{1}{2}\min_{\ket{\psi}_{AA'BB'}}(S_{AB}+S_{AA'}+S_{BB'}-S_{A'}-S_{B'})\\
 & \equiv\min_{\ket{\psi}_{AA'BB'}}f^{R}(A,A',B,B').
\end{align}
The $Q$-correlation and the $R$-correlation are non-increasing under
local operations, but not necessarily under LOCC. They satisfy the
inequality \citep{LevinSmith19:OptimizedMeasures}
\begin{equation}
\frac{1}{2}I\leq E_{Q},E_{R}\leq E_{P}.\label{eq:IQRPineq}
\end{equation}

We note a close relationship between the $R$-correlation and the
CEMI, which is clear from the following expression of $E_{R}$
\begin{align}
E_{R}(A:B) & =\frac{1}{2}\min_{\ket{\psi}_{AA'BB'}}(I(AA':BB')-I(A':B')).\label{eq:ERandCEMI}
\end{align}
It is similar to the CEMI (\ref{eq:CEMI}), though the minimization
of the CEMI is performed over all possible extensions.

\subsubsection*{The Holographic Counterparts}

Here we investigate holographic duals of the $Q$-correlation and
the $R$-correlation. The definition of the holographic dual candidate
of $E_{Q}$ is stated as follows (we focus on static geometries):

Given an entanglement wedge $\mathcal{M}_{AB}$, divide its boundary
into $\partial\mathcal{M}_{AB}=\mathcal{A}\cup\mathcal{B}$ so that
$\mathcal{A}=A\cup A'$ and $\mathcal{B}=B\cup B'$. Then minimize
the combination of holographic entanglement entropy $f^{Q}(A,A',B,B')$
over all possible partitions. We define the minimum as the \textit{entanglement
wedge mutual information} (EWMI), denoted by $E_{M}$
\begin{figure}
\includegraphics[scale=0.2]{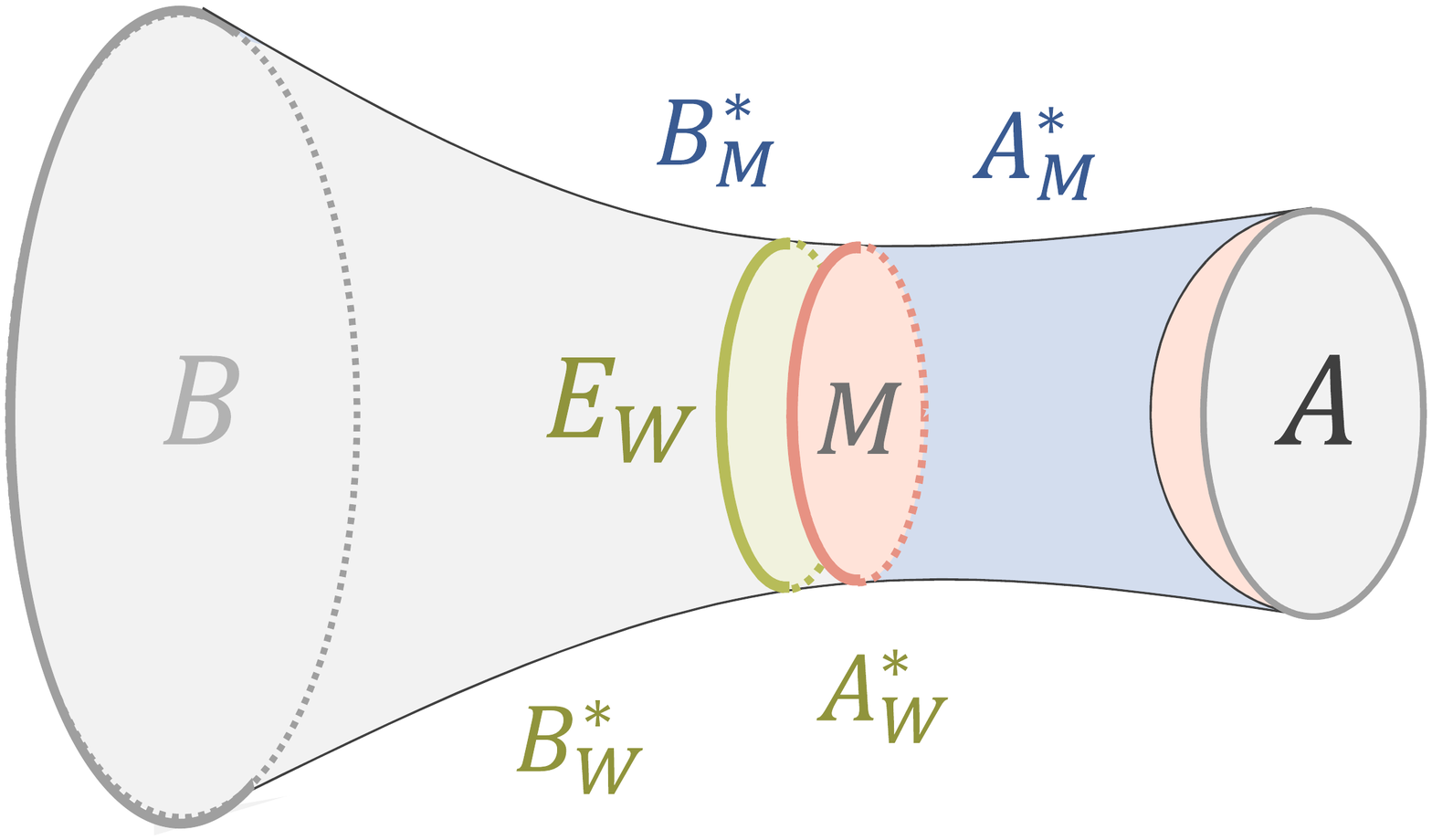}
\caption{\label{fig:EM} The entanglement wedge mutual information $E_{M}$
in the entanglement wedge. In the above picture, $E_{M}$ is given
by the area of red codimension-2 surfaces subtracted by the area of
blue codimension-2 surface (divided by $2\cdot4G_{N}$), which may
be understood as the mutual information $\frac{1}{2}I(A:M)$. The
symmetry $E_{M}(A:B)=E_{M}(B:A)$ stems from the fact that the RT-surface
of $S_{BA'}$ and $S_{AB'}$ have the same configurations. The optimal
partition $A_{M}^{*}$ and $B_{M}^{*}$ of the EWMI located on the
RT-surface of $S_{AB}$, are not necessarily equivalent to these $A_{W}^{*}$
and $B_{W}^{*}$ of the EWCS.}
\end{figure}
\begin{align}
E_{M}(A:B) & :=\min_{A'\cup B'}f^{Q}(A,A',B,B').
\end{align}
An example of the EWMI is depicted in the Fig.\ref{fig:EM}. It may
be regarded as the half of the mutual information between $A$ (or
$B$) and the ``subsystem'' $M$ assigned to the codimention-2 cross
section of $S_{AA'}\ (=S_{BB'})$.

Generically, the EWMI requires us to consider many complicated configurations of $A'$ and $B'$ in order to minimize $f^Q$. For some simple cases, however, such as the two disjoint intervals in $\rm{AdS}_3/\rm{CFT}_2$ or the (symmetric) Araki-Lieb saturating configurations, there is an intuitive way to compute $E_{M}$ owing to the symmetry of setup: Minimize (half
of) the mutual information $\max\{I(A:M),I(B:M)\}$ over all possible
choices of the cross sections, 
\begin{equation}
E_{M}(A:B)=\frac{1}{2}\min_{M}(\max\{I(A:M),\ I(B:M)\}),\label{eq:EMminmax}
\end{equation}
where $M$ corresponds to the cross section of some partition $A'\cup B'$.
This form also clarifies a useful relation 
\begin{equation}
I(A:M^{*})=I(B:M^{*}),\label{eq:IAMIBM}
\end{equation}
for at least one of the optimal cross sections $M^{*}$. Note that
the optimal purification for $E_{M}$ is not necessarily unique, nor
does it necessarily agree with that of $E_{W}$ (Fig.\ref{fig:EM}).
We will see both concrete examples in the below discussion of the
Araki-Lieb transition of $E_{M}$. There is another suggestive form of $E_{M}$ for these cases, 
\begin{align}
 & E_{M}(A:B)\nonumber \\
 & =\frac{1}{2}\left[\frac{1}{2}I(A:B)\right.\nonumber \\
 & \left.+\min_{A'\cup B'}\left(S_{AA'}+\frac{I(A:B')+I(B:A')}{2}\right)\right],
\end{align}
where we have used $I(A':B')=0$ which holds for the ancillary subsystems
on the RT-surface. At least one of $I(A:B')$ and $I(B:A')$ must
vanish at this point because the whole system is homologically trivial.
Moreover, the balancing condition (\ref{eq:IAMIBM}) is equivalent
to the condition $I(A:B^{*})=I(B:A^{*})$. Thus we can conclude that
both of $I(A:B')$ and $I(B:A')$ should vanish for the balanced optimal
partition. As a result, we reach a formula
\begin{equation}
E_{M}(A:B)=\frac{1}{2}\left[\frac{1}{2}I(A:B)+S_{AA_{b}^{*}}\right],\label{eq:EM_ISAAstar}
\end{equation}
where $A_{b}^{*}$ is the balanced optimal partition. We may define
the deviation from the EWCS due to the balancing term $S_{BA'}$ as
\begin{equation}
D_{b}(A:B):=S_{AA_{b}^{*}}-E_{W}(A:B)\geq0.\label{eq:Dev}
\end{equation}
We will check this formula (\ref{eq:EM_ISAAstar}) by direct computation
in the Araki-Lieb transition.

%%Added for v2
A caveat is that neither the formula (\ref{eq:EMminmax}) nor (\ref{eq:EM_ISAAstar}) is necessarily valid for any configurations, and there possibly exists other types of optimal configurations of $A'$ and $B'$ for more complicated subsystems. Indeed, for example, if we set $|B_1|>|B_2|$ in the Araki-Lieb saturating configuration, $E_M$ can be realized by an optimal configuration neither of $I(A:M^*)$ nor $I(B:M^*)$, but of a combination $I(A:M^*_A)+I(B:M^*_B)$ where $M^*_A\cup M^*_B=M$. Such a configuration is not preferred for the disjoint two intervals \citep{LevinSmith19:Toappear} nor for the symmetric Araki-Lieb configuration. This example indicates that we need to replace $\max\{I(A:M), I(B:M)\}$ in (\ref{eq:EMminmax}) with $\max_{M_A \cup M_B =M}\{I(A:M_A)+I(B:M_B)\}$ in general. We leave proving or disproving it for generic configurations as an important future work.
%%%%%%%%%%%%

The EWMI satisfies the properties of $E_{Q}$. For example, it can
not be greater than the EWCS,
\begin{equation}
E_{M}\leq E_{W},
\end{equation}
which must hold to be consistent with $E_{W}=E_{P}$ from (\ref{eq:IQRPineq}).
One can prove this inequality by drawing a picture, but an easier
way is to use the von Neumann entropy to represent the corresponding
geometrical areas. Suppose the optimal partition of $E_{W}$ is given
by $A_{W}^{*}$ and $B_{W}^{*}$. Then we can show $E_{W}=S_{AA_{W}^{*}}\geq\frac{1}{2}(S_{A}+S_{B}+S_{AA_{W}^{*}}-S_{BA_{W}^{*}})\geq E_{M}$,
where we have used strong subadditivity. 

\begin{figure}
\includegraphics[scale=0.2]{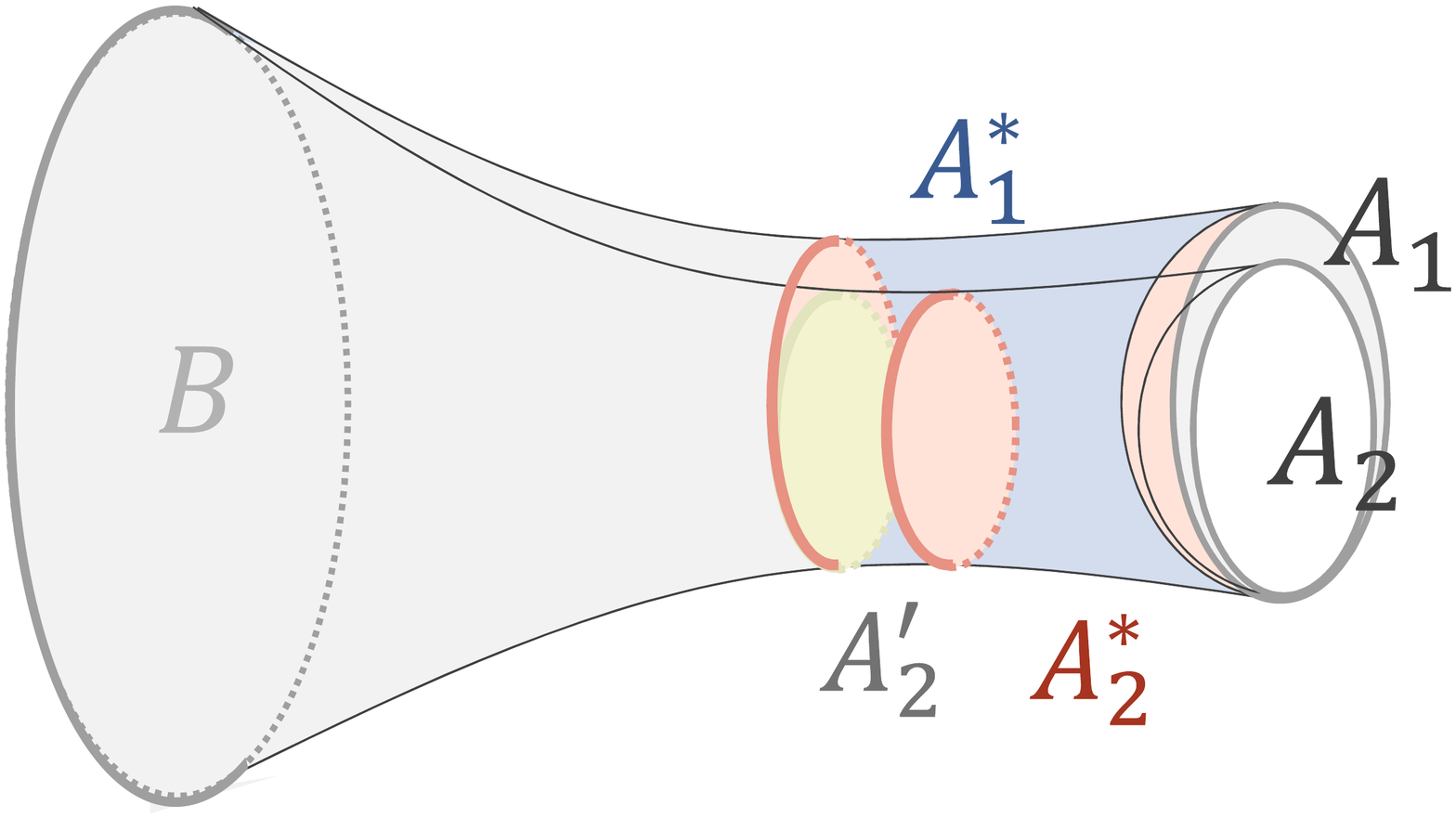}
\caption{\label{fig:EMextensivity} The extensivity $E_{M}(A_{1}:B)\protect\geq E_{M}(A_{2}:B)$
for $A_{1}\supset A_{2}$. We abbreviate the $B'$ labels. From the
optimal partition $A_{1}^{*}$ for $A_{1}$, one can induce a partition
$A'_{2}$ on $\partial\mathcal{M}_{A_{2}B}$ so that $A_{1}^{*}\cap\gamma_{A_{2}B}=A'_{2}\cap\gamma_{A_{2}B}$.
Then $E_{M}(A_{1}:B)\protect\geq f^{Q}(A_{2},B,A_{2}')$ holds due
to the minimality of RT-surface, and $f^{Q}(A_{2},B,A_{2}')\protect\geq E_{M}(A_{2}:B)$
is clear by definition.}
\end{figure}

Similarly, $E_{M}$ can not be less than half of holographic mutual
information,
\begin{equation}
\frac{1}{2}I\leq E_{M}.
\end{equation}
It is clear from (\ref{eq:EM_ISAAstar}) as $S_{AA_{b}^{*}}\geq E_{W}(A:B)\geq\frac{1}{2}I(A:B)$.
These properties also guarantee that $E_{M}(A:B)=S_{A}=S_{B}$ for
pure states, and that $E_{M}$ vanishes if and only if $I(A:B)=0$
(with $A^{*}=\gamma_{A}$ and $B^{*}=\gamma_{B}$). It also shows
the extensivity $E_{M}(A_{1}:B)\geq E_{M}(A_{2}:B)$ when $A_{1}\supset A_{2}$
(Fig.\ref{fig:EMextensivity}). The additivity $E_{M}(\rho_{A_{1}B_{1}}\otimes\sigma_{A_{2}B_{2}})=E_{M}(\rho_{A_{1}B_{1}})+E_{M}(\sigma_{A_{2}B_{2}})$
is also clear because the decoupled state corresponds to disjoint
geometries. All of these consistent properties tempt us to propose
the relation (at the leading order $O(N^{2})$) 
\begin{equation}
E_{Q}=E_{M}.
\end{equation}

\begin{figure}
\includegraphics[scale=0.14]{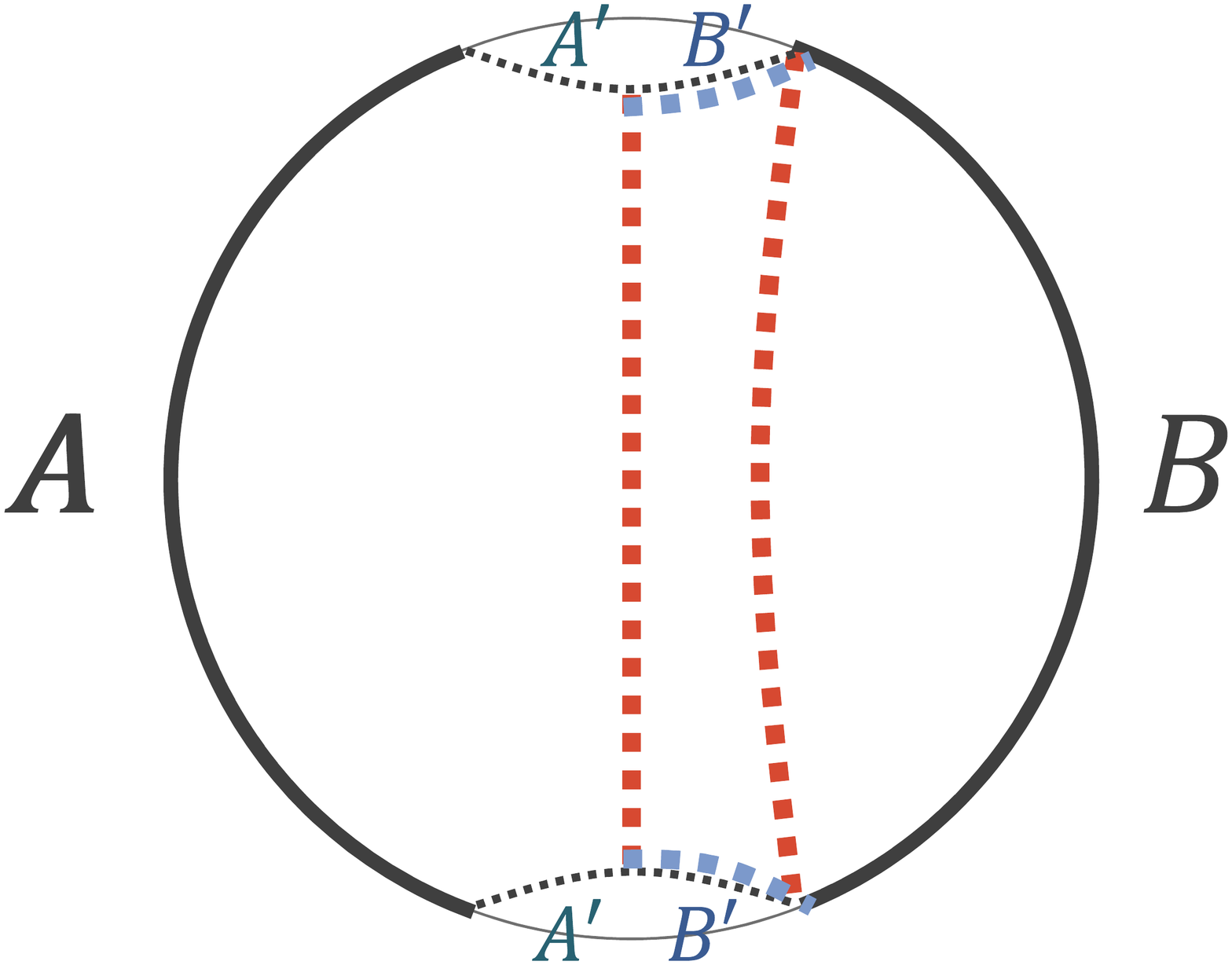}
\includegraphics[scale=0.14]{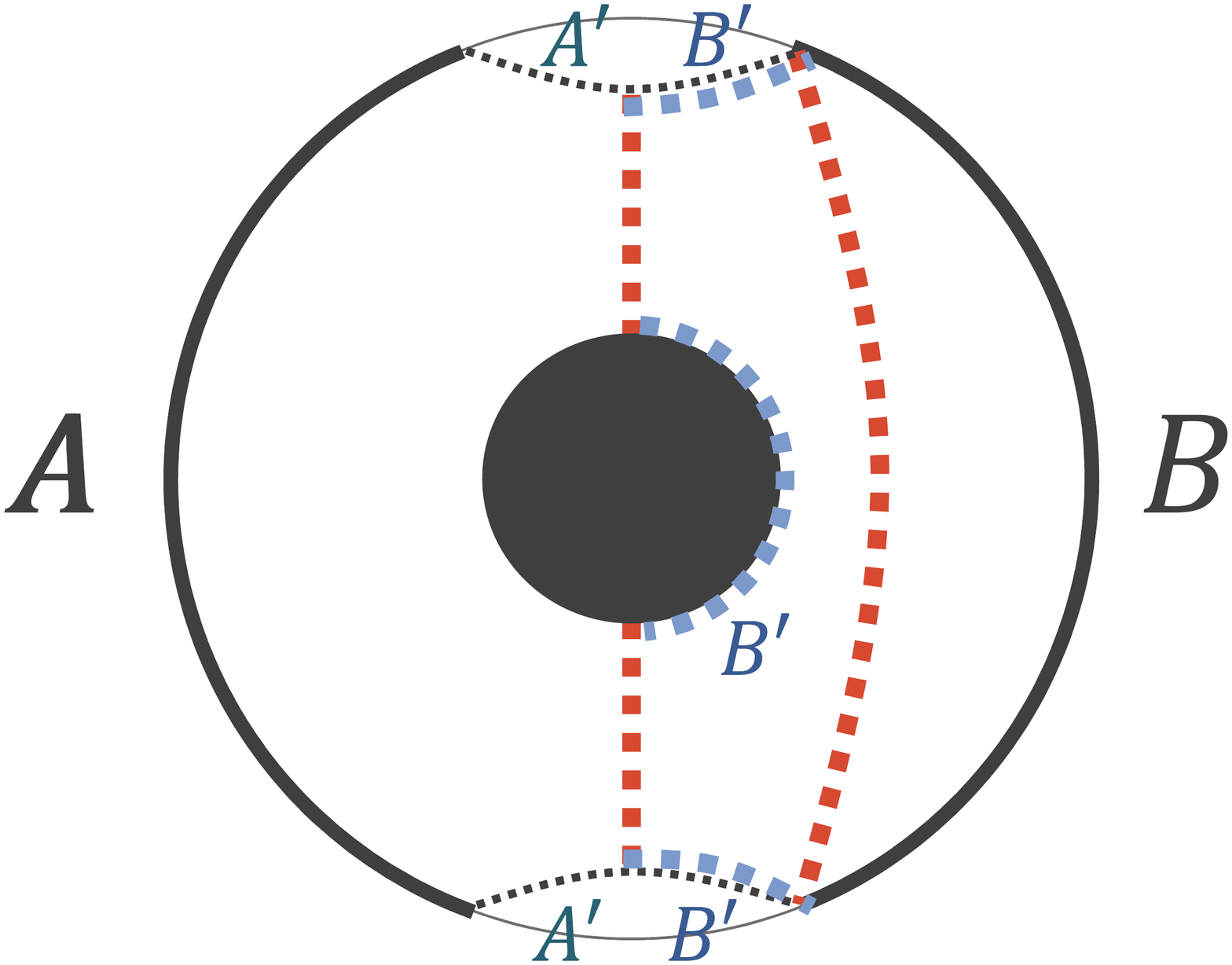}
\caption{\label{fig:EMin3D} The $E_{M}$ in the pure global ${\rm AdS}_{3}$
(Left) and in the global BTZ (Right) for the symmetric two disjoint
intervals, in which $E_{M}=\frac{1}{2}(\frac{1}{2}I+E_{W})$. In the
vacuum, one can map the two disjoint subsystems into this setup by
the conformal symmetry.}
\end{figure}

In pure ${\rm AdS}_{3}$, the $E_{M}$ for two disjoint intervals
has a simple expression (Fig.\ref{fig:EMin3D}).
In such cases, the optimal partition coincides with that of $E_{W}$,
as it is obvious from the conformal symmetry. Thus $E_{M}$ becomes
just the average of $\frac{1}{2}I$ and $E_{W}$ by (\ref{eq:EM_ISAAstar}),
\begin{equation}
E_{M}(A:B)=\frac{1}{2}\left[\frac{1}{2}I(A:B)+E_{W}(A:B)\right].
\end{equation}
From this expression, we can easily confirm all of the properties
of $E_{M}$ mentioned above. This expression is not necessarily true
in generic setups such as three or more multipartite intervals or
black hole geometry. 

\begin{figure}
\includegraphics[viewport=55mm 0bp 785bp 595bp,scale=0.12]{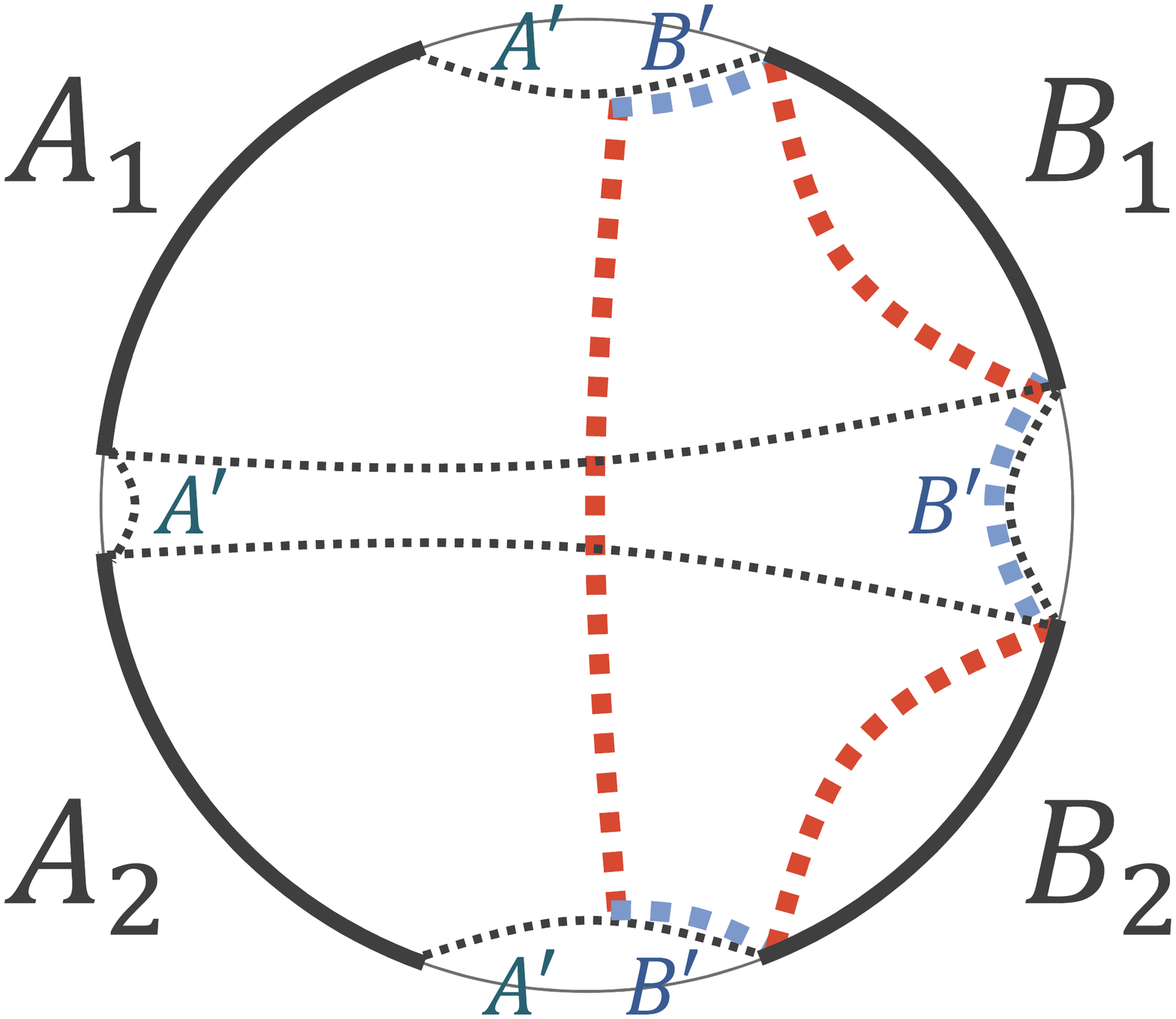}
\includegraphics[viewport=55bp 0bp 785bp 595bp,scale=0.12]{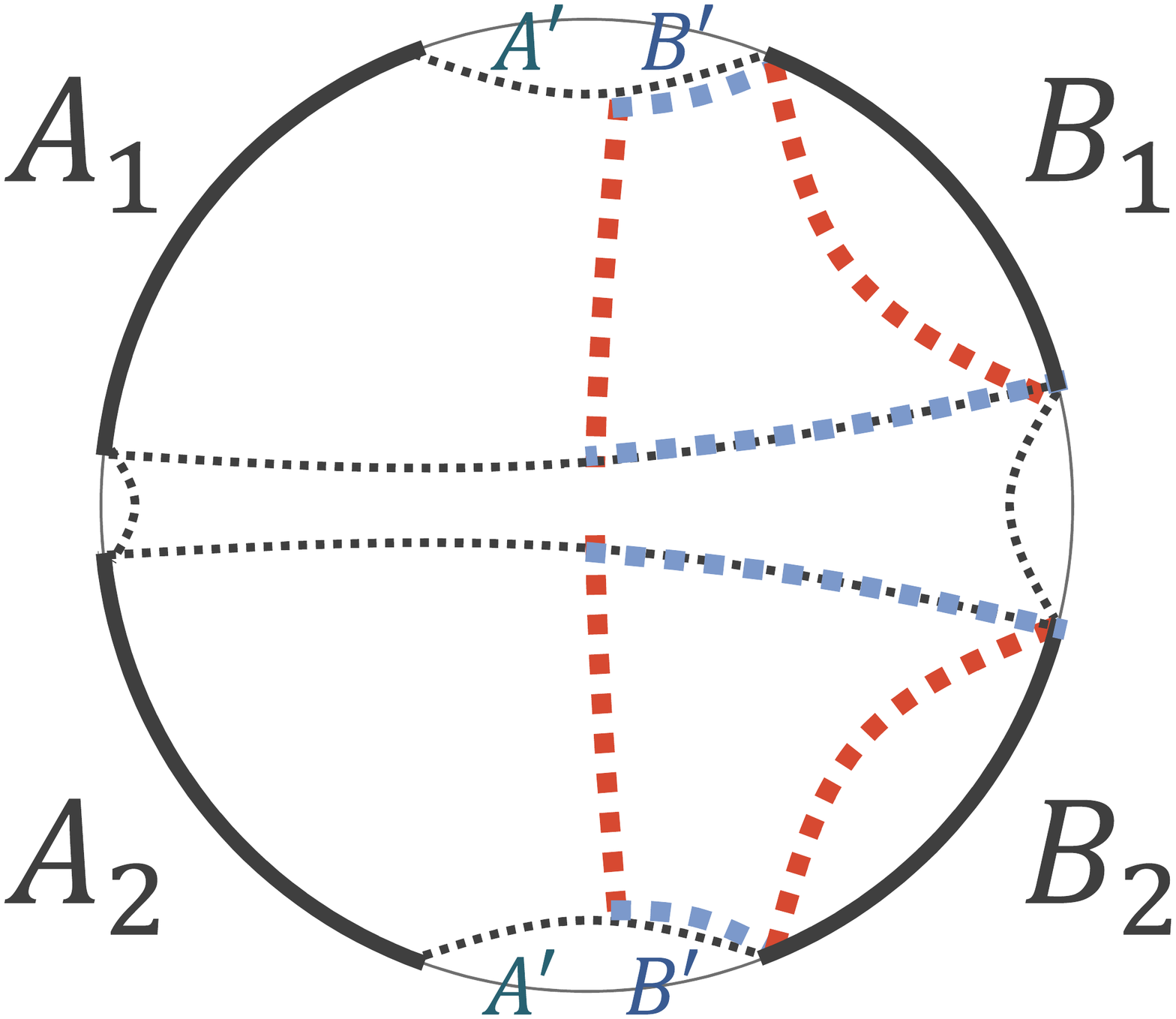}
\includegraphics[viewport=55bp 0bp 785bp 595bp,scale=0.12]{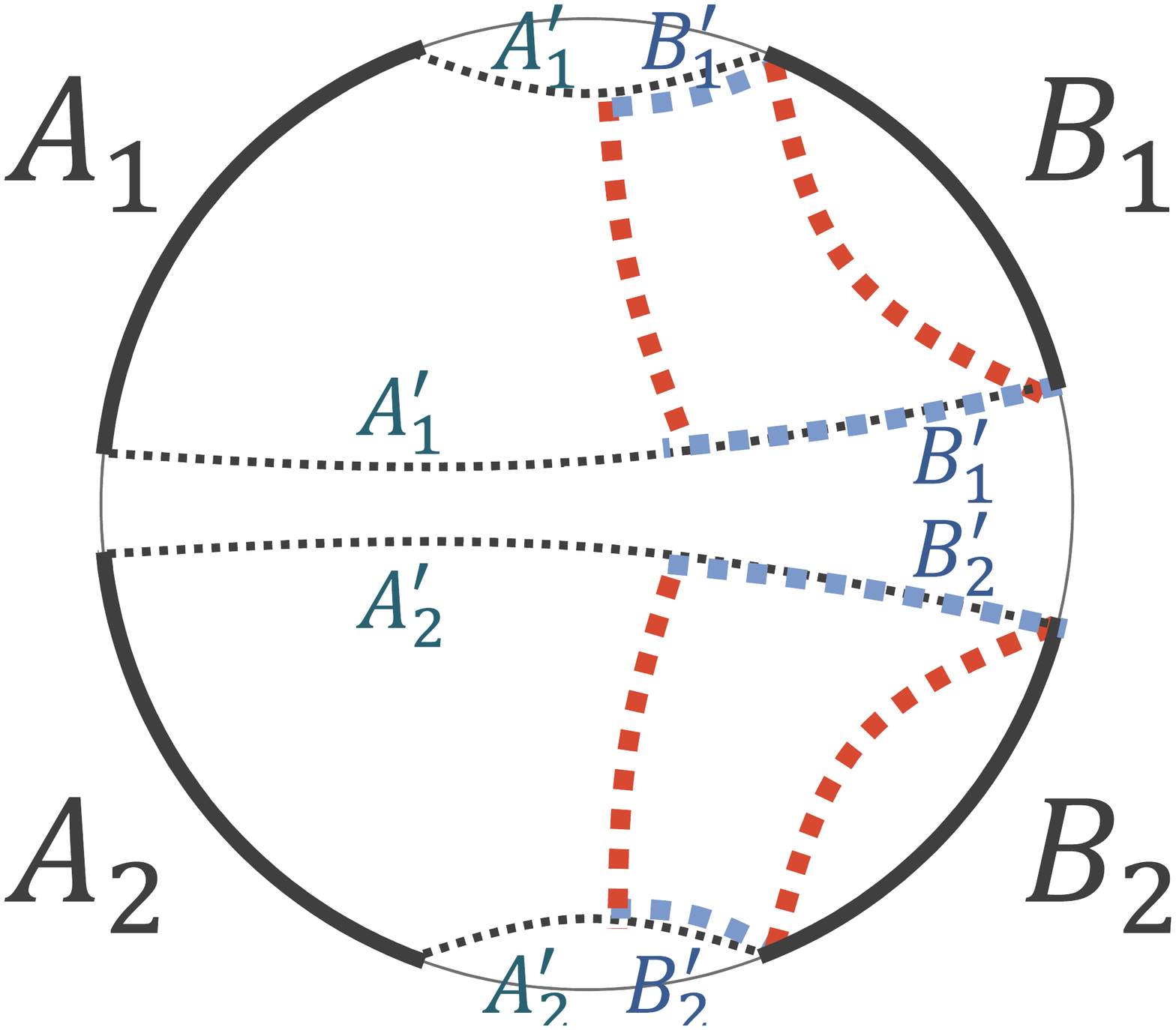}
\caption{\label{fig:EM_SSA} A proof of the strong superadditivity for $E_{M}$.
The left figure corresponds to $E_{M}(\rho_{A_{1}A_{2}B_{1}B_{2}})$,
and the right one does to $E_{M}(\rho_{A_{1}B_{1}})+E_{M}(\rho_{A_{2}B_{2}})$.
As the total of the areas, (Left) $\protect\geq$ (Middle) $\protect\geq$
(Right) is obvious (plus sign for red and minus sign for blue).}
\end{figure}

Surprisingly, the EWMI also satisfies the strong superadditivity 
\begin{equation}
E_{M}(\rho_{A_{1}A_{2}B_{1}B_{2}})\geq E_{M}(\rho_{A_{1}B_{1}})+E_{M}(\rho_{A_{2}B_{2}}),\label{eq:EM_SSA}
\end{equation}
which can be proven geometrically (Fig.\ref{fig:EM_SSA}). It is similar
to the proof of the strong superadditivity of $E_{W}$ \citep{UmemotoTakayanagi18:Entanglementofpurificationthroughholographicduality}.
The relation (\ref{eq:EM_SSA}) is not a generic property of $E_{Q}$. Thus we may regard it as a characteristic of holographic correlations,
as with the holographic entropy cone \citep{HaydenHeadrickMaloney11:HolographicMutualInformation,BaoNezamiOoguriStoicaSullyWalter15:TheHolographicEntropyCone,RotaWeinberg18:Newconstraintsforholographicentropyfrommaximin,HubenyRangamaniRota18:TheHolographicentropyArrangement,Hernandez19:HolographicentropyconeforFive}. 

The dual of $E_{R}$ is defined in the same manner, replacing $f^{Q}$
with $f^{R}$ in the above procedure. However, it turns out that this
definition is equivalent to that of the EWCS. It stems from the fact
we implicitly used in the definition of $E_{M}$ (and $E_{W}$) that
it is sufficient to consider the ancillary systems $A'$ and $B'$
located only on the RT-surface $S_{AB}$ for minimization. For such
subsystems we find $I(A':B')=S_{A'}+S_{B'}-S_{AB}=0$, resulting in
$E_{R}=E_{P}=E_{W}$ from (\ref{eq:ERandCEMI}). We also state it
as a holographic proposal
\begin{equation}
E_{R}=E_{W}.
\end{equation}
The additivity of the EWCS is consistent with that of the $R$-correlation
\citep{LevinSmith19:OptimizedMeasures}. The relation between $E_{R}$
and $E_{I}$ then gives an interesting perspective on the geometrical
extensions: if only pure geometries are available, the correlations
can reduce to $E_{W}$ at most. If mixed geometries are also allowed,
then the inaction gives a further reduction to $\frac{1}{2}I$. 

\subsection{The EWMI in the Araki-Lieb Transition}

Let us study $E_{M}$ in the Araki-Lieb transition discussed in section
\ref{subsec:EWCS_ALtransition} in detail. First, we remark that $E_{M}=S_{A}$
should hold for $p<p_{{\rm MI}}^{*}$ from the inequality $\frac{1}{2}I\leq E_{M}\leq E_{W}$,
while it also can be checked by direct computation. For $p>p_{{\rm MI}}^{*}$,
the situation is more complicated than $E_{W}$ due to the four configurations
of $S_{AA'}-S_{BA'}$. For simplicity, we fix the size $b$ to unit
size in the setup and always deal with the relative size $p$ as the
parameter. 

\begin{figure}
\includegraphics[scale=0.14]{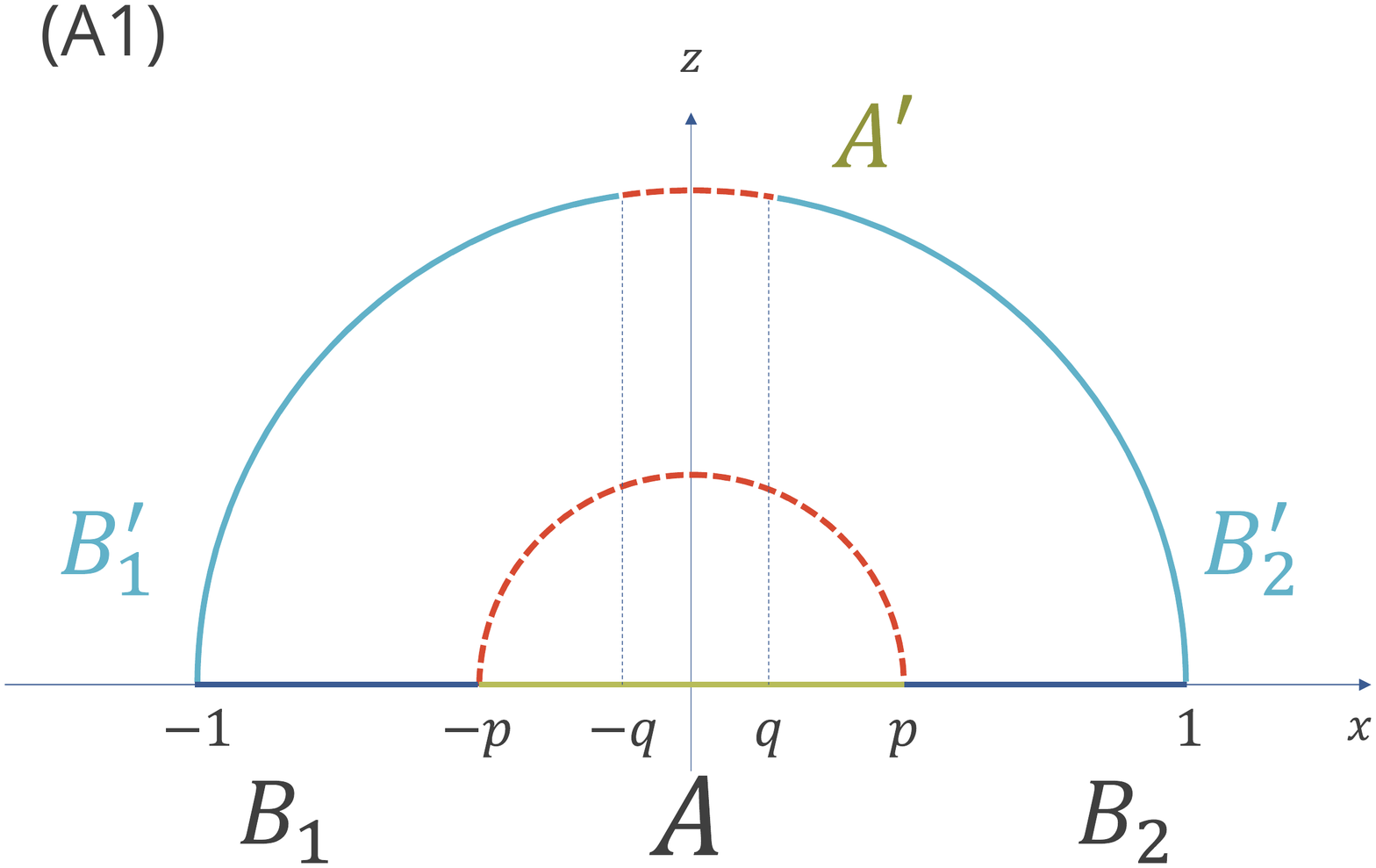}
\includegraphics[scale=0.14]{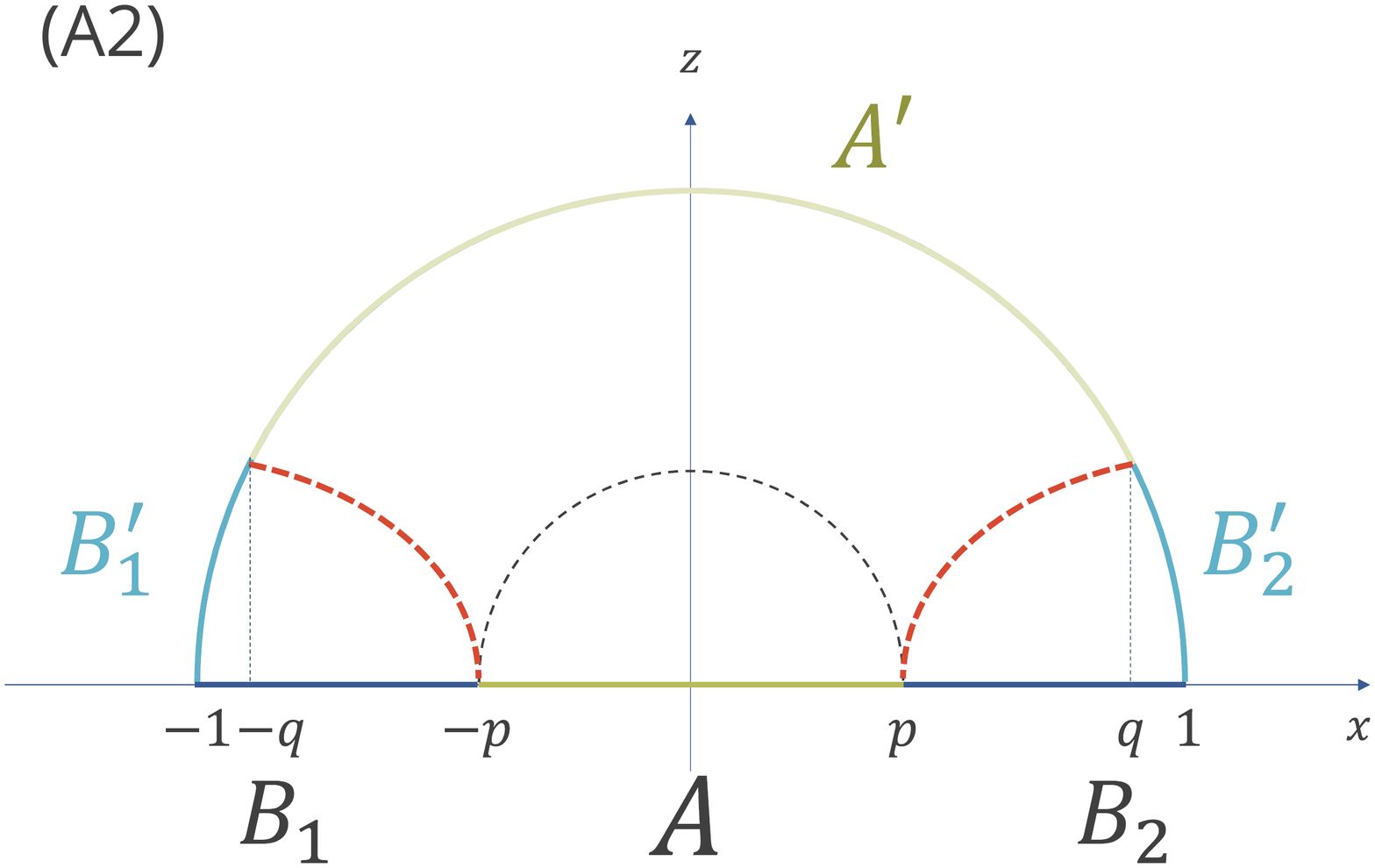}
\includegraphics[scale=0.14]{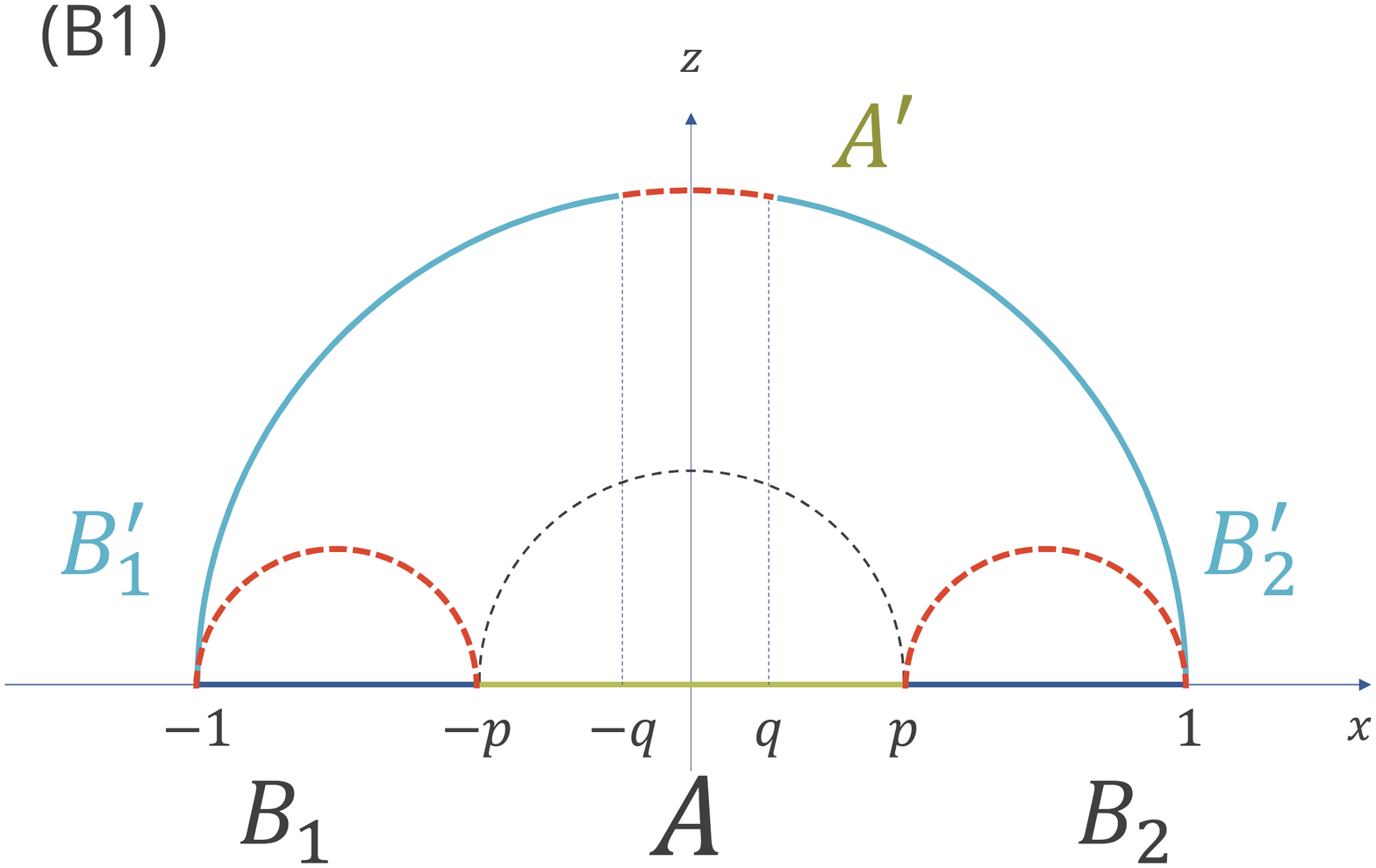}
\includegraphics[scale=0.14]{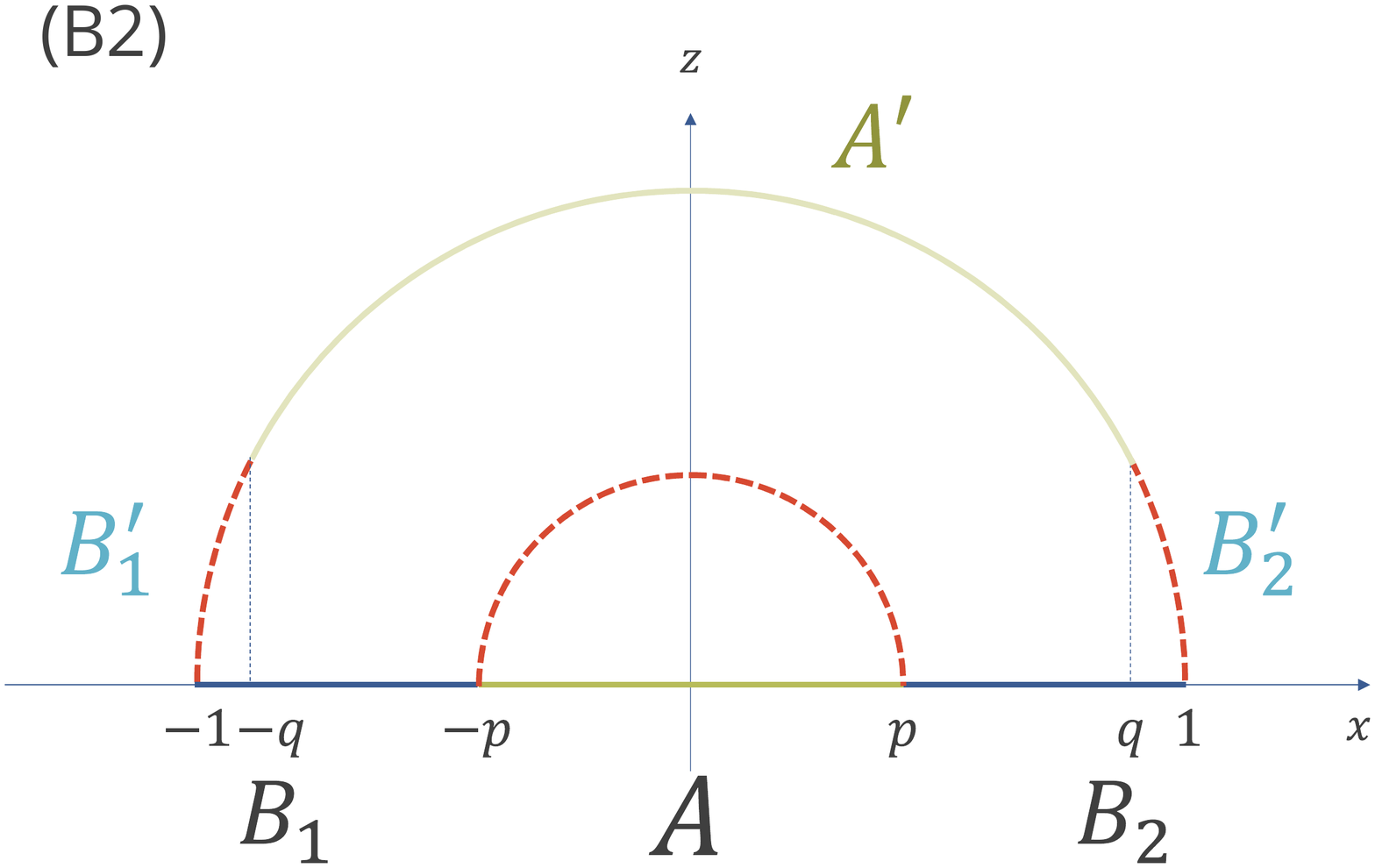}
\caption{\label{fig:EM_ALtransition} The two phases of $S_{AA'}$ (top panels)
and these of $S_{BA'}$ (bottom panels). The RT-surface of $S_{AA'}$
and $S_{BA'}$ are denoted as red dashed lines. }
\end{figure}

The two phases of $S_{AA'}$ and the two phases of $S_{BA'}$ are
depicted in Fig.\ref{fig:EM_ALtransition}. 
 The minimal configuration depends not only on the parameter $p$
but also on the size of $A'$, parameterized by $q\in(0,1)$. We can
easily find out the minimal configurations in the extremal cases:
in the small $A'$ limit ($q\to0$), the phase (A1) for $S_{AA'}$
and the phase (B1) for $S_{BA'}$ are preferred (recall $I(B_{1}:B_{2})=0$
for $p>p_{{\rm MI}}^{*}$). Similarly, we have the phase (A2) for
$S_{AA'}$ and the phase (B2) for $S_{BA'}$ in the large $A'$ limit
($q\to1$). Therefore, as we increase $q$ from $0$ to $1$, we will
see a phase transitions of $S_{AA'}-S_{BA'}$ for the fixed $p>p_{{\rm MI}}^{*}$
in either path
\begin{align}
{\rm (I)\ (A1,B1)\to(A1,B2)\to(A2,B2)},\\
{\rm (II)\ (A1,B1)\to(A2,B1)\to(A2,B2)}.
\end{align}
Note that $S_{BA'}$ is always in (B2) regardless to $q$ for $p<p_{{\rm MI}}^{*}$.

It is not hard to show that increasing $q$ may decrease $S_{AA'}-S_{BA'}$
only in the phase (A2, B1). In the phase (A1, B1), changing $q$ has
no effect at all, and in the phase (A1, B2) and (A2, B2), increasing
$q$ does increase $S_{AA'}-S_{BA'}$. Therefore, a nontrivial optimal
partition for $E_{M}$ is observed only when the phase transition
follows the path (II). 

\begin{figure}
\includegraphics[scale=0.29]{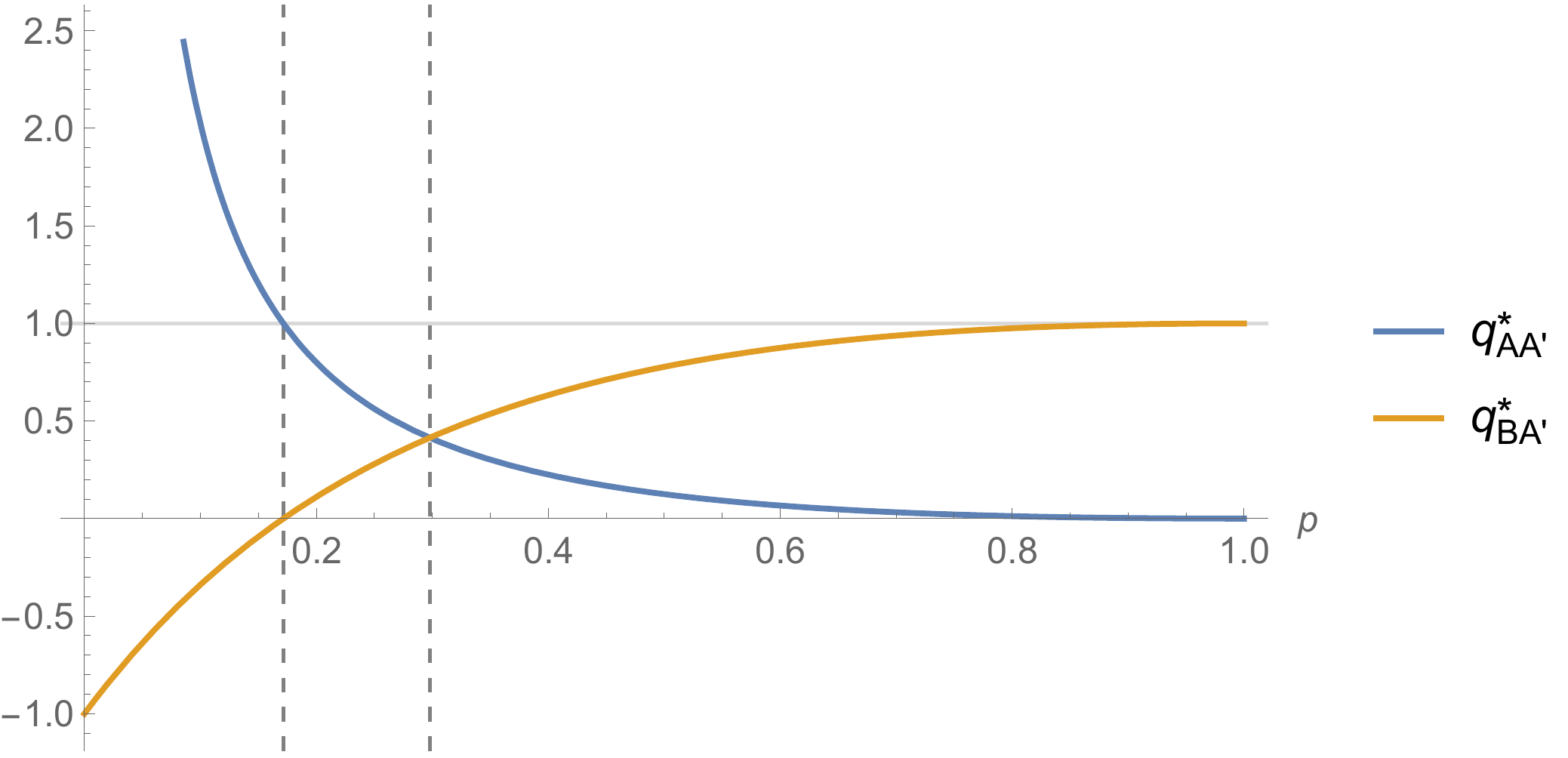}
\caption{\label{fig:qstars_plot} The phase transition points of $S_{AA'}$
(blue) and $S_{BA'}$ (yellow) w.r.t. the size $q$. The vertical
dashed lines denote $p_{{\rm MI}}^{*}=3-2\sqrt{2}\simeq0.17$ and
$p_{{\rm EM}}^{*}=-1+2\sqrt{2}-2\sqrt{2-\sqrt{2}}\simeq0.30$.}
\end{figure}

With this in mind, we find the phase transition points $q^{*}$ of
$S_{AA'}$ and $S_{BA'}$ as a function of $p$
\begin{equation}
q_{AA'}^{*}(p)=\frac{(1-p)^{2}}{4p},
\end{equation}
\begin{equation}
q_{BA'}^{*}(p)=-\frac{1-6p+p^{2}}{(1+p)^{2}}.
\end{equation}
These are plotted in Fig. \ref{fig:qstars_plot}.
 The region $q<q_{AA'}^{*}$ corresponds to the phase (A1) for $S_{AA'}$,
and the region $q<q_{BA'}^{*}$ to the phase (B1) for $S_{BA'}$.
The phase transition of $E_{M}$ happens at the crossing point $p_{{\rm EM}}^{*}$
at which $q_{AA'}^{*}(p_{{\rm EM}}^{*})=q_{BA'}^{*}(p_{{\rm EM}}^{*})$
holds,
\begin{equation}
p_{{\rm EM}}^{*}=-1+2\sqrt{2}-2\sqrt{2-\sqrt{2}}\simeq0.30.
\end{equation}
The ancillary system $A'$ of \textit{any} size $q\leq q_{BA'}^{*}$
achieves the minimum $E_{M}=S_{A}$ for $p<p_{{\rm EM}}^{*}$. For
$p>p_{{\rm EM}}^{*}$, the minimum of $S_{AA'}-S_{BA'}$ is obtained
at $q=q_{BA'}^{*}(p)$. 

Therefore, we have found $E_{M}=S_{A}$ for $p\leq p_{{\rm EM}}^{*}$,
and $E_{M}=\frac{1}{2}(S_{A}+S_{B}+S_{AA^{*}}-S_{BA^{*}})=\frac{1}{2}(S_{A}-S_{A^{*}}+S_{AA^{*}})$
for $p>p_{{\rm EM}}^{*}$. In the latter case, the size of $A^{*}$
is given by $q_{BA'}^{*}(p)$, and $S_{AA^{*}}$ is in the phase (A2)
and $S_{BA^{*}}$ is at the phase transition point (B1)$=$(B2). We
may compute $S_{A^{*}}$ as $S_{A^{*}}=\frac{1}{2}(S_{A}+S_{AB}-S_{B})$
from the equality condition (B1)=(B2).

\begin{figure}
\includegraphics[scale=0.3]{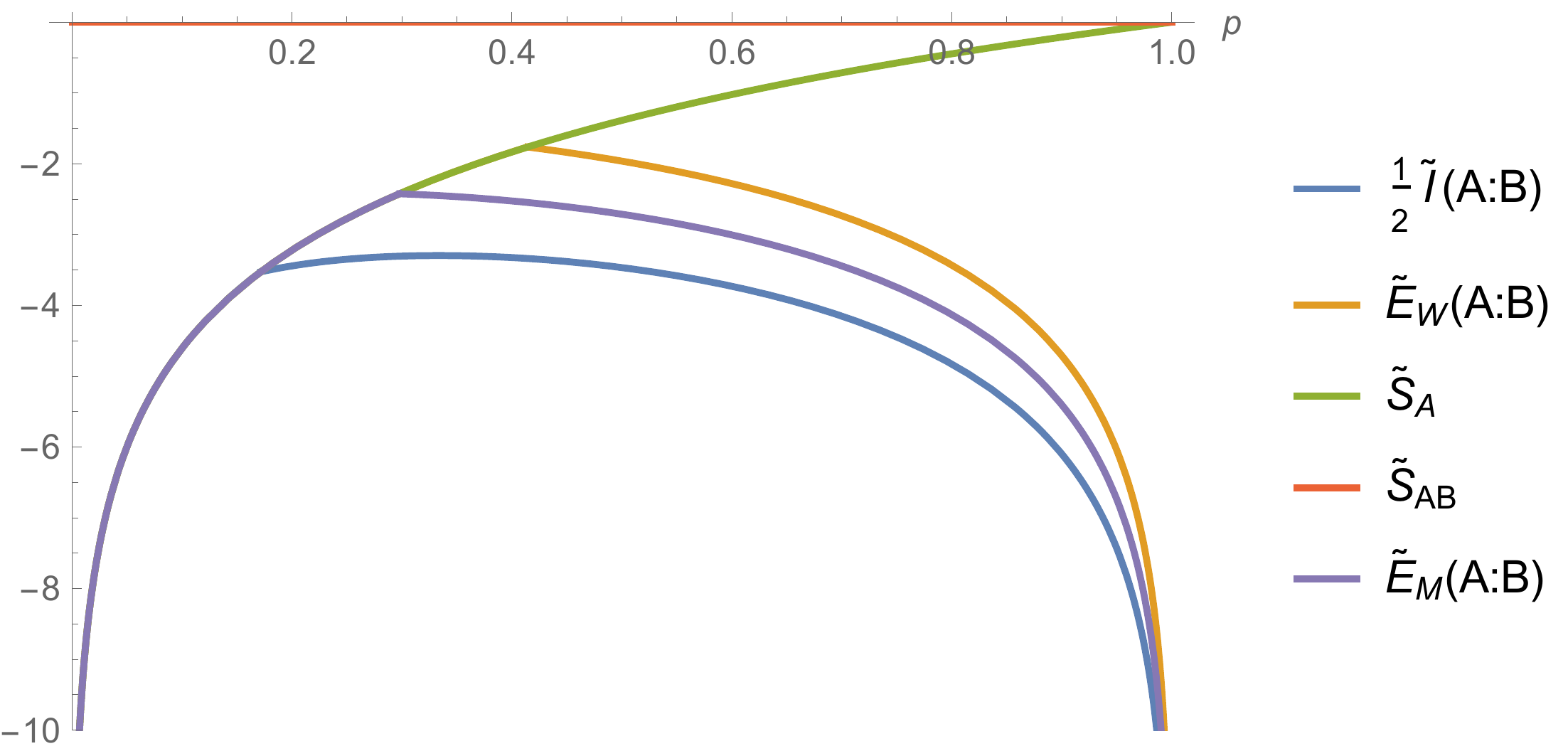}
\caption{\label{fig:EMforALtransition} The $E_{M}$ for the Araki-Lieb transition
(normalized by subtracting $S_{AB}$). }
\end{figure}

\begin{figure}
\includegraphics[scale=0.32]{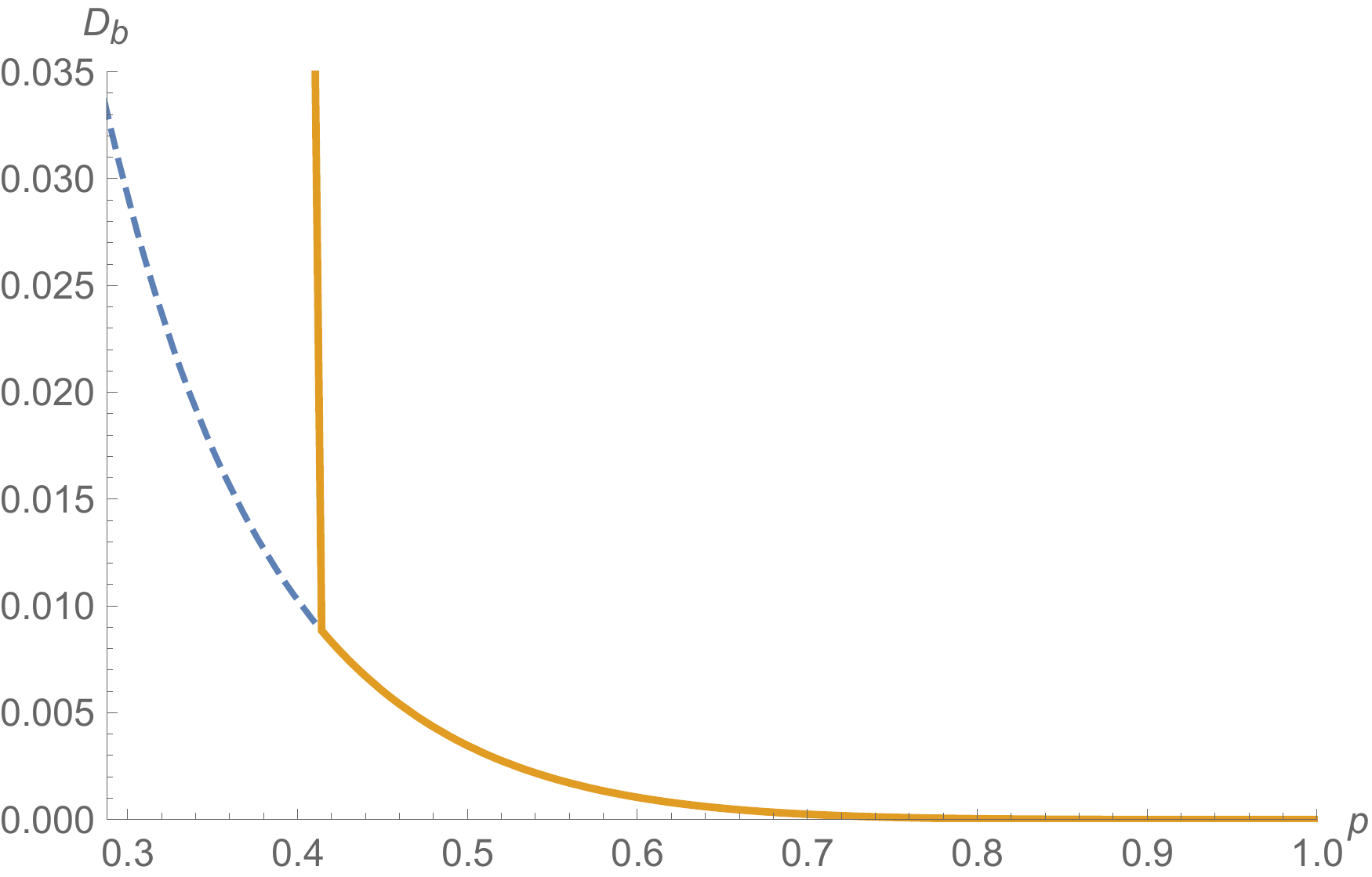}
\caption{\label{fig:Db} The deviation $D_{b}$ for the Araki-Lieb transition
(yellow solid line), and the $D_{b}$ replaced $E_{W}$ with the non-optimal
configuration in the Fig. \ref{fig:EWCSALtransition} (blue dashed
line). }
\end{figure}

After all, we obtain $E_{M}$ in the Araki-Lieb transition as
\begin{align}
 & E_{M}(A:B)\nonumber \\
 & =\begin{cases}
S_{A} & (p<p_{{\rm EM}}^{*})\\
\frac{1}{2}\left[\frac{1}{2}I(A:B)+S_{AA^{*}}(p,q_{BA'}^{*}(p))\right] & (p>p_{{\rm EM}}^{*})
\end{cases},\label{eq:EM_ALtransition}
\end{align}
where $S_{AA^{*}}\equiv S_{AA^{*}}(p,q_{BA'}^{*}(p))$ denotes a contribution
from the geodesics between $\partial A$ and $\partial A^{*}$, 
\begin{equation}
S_{AA^{*}}(p,q_{BA'}^{*}(p))=\frac{c}{3}\log\left(\frac{(1-p)(1+6p+p^{2})}{4\sqrt{p}\epsilon}\right).
\end{equation}
This result (\ref{eq:EM_ALtransition}) confirms the shortcut formula
(\ref{eq:EM_ISAAstar}). Note that $\frac{1}{2}I(A:B)=S_{A}$ for
$p<p_{{\rm MI}}^{*}$ and that the balanced optimal partition for
$p\in(p_{{\rm MI}}^{*},p_{{\rm EM}}^{*})$ is given by $A'$ of the
size $q=q_{BA'}^{*}(p)$, not the trivial partition (though it is
also optimal). The balancing condition $I(A:M^{*})=I(B:M^{*})$ generically
corresponds to the condition (B1)$=$(B2). The deviation (\ref{eq:Dev})
is given by
\begin{equation}
D_{b}(p)=S_{AA^{*}}(p,q_{BA^{'}}^{*})-E_{W}(p)=\frac{c}{3}\log\frac{1+6p+p^{2}}{4\sqrt{p}(1+p)}.
\end{equation}
The plots of $E_{M}$ and $D_{b}$ are given in Fig.\ref{fig:EMforALtransition}
and Fig.\ref{fig:Db}.

In particular, the strict inequality $p_{{\rm MI}}^{*}<p_{{\rm EM}}^{*}$
indicates that $E_{M}$ must be strictly greater than the axiomatic
entanglement measures for $p\in(p_{{\rm MI}}^{*},p_{{\rm EM}}^{*})$,
based on the same logic as the EWCS. One can also confirm that $E_{M}$
exhibits the same kind of phase transition in the global BTZ black
hole.

\section{Discussion}

We have introduced a series of possible holographic duals to the optimized
correlation measures. The crucial assumption for the equivalence was
that the geometrical extensions are enough to achieve their minimum
in holographic CFTs. They demonstrate many properties which are completely
consistent with the original information-theoretic measures. 

We showed that the EWCS and the EWMI can be larger than the wide class
of entanglement measures, while the holographic mutual information
is not necessarily. This implies that the EWCS and the EWMI should
be more sensitive to classical correlations than the holographic mutual
information. Note that both of the EWCS and the EWMI satisfy the strong
superaddivity, which is a weaker property of quantum entanglement
than the monogamy relation (since the latter induces the former).
In addition, the EOP or the reflected entropy is supposed to be more
sensitive to classical correlations than the mutual information \citep{BhattacharyyaJahnTakayanagiUmemoto19:EntanglementofPurificationinManybody,KusukiTamaoka19:DynamicsofEntanglementWedge}.
It will be interesting future work to investigate a role of classical
correlation in holographic CFTs.

There is a caveat that all of our discussions are restricted to the
leading order $O(N^{2})$. In particular, the Araki-Lieb transition
relies on the property of the holographic entanglement entropy at
this order. If one includes quantum corrections from bulk entanglement
entropy at $O(N^{0})$ \citep{FaulknerLewkowyczMaldacena13:QuantumcorrectiontoHolographic,BarrellaDongHartnollMartin13:HolographicEntanglementBeyondClassical},
the rigorous relation will be violated. For instance, the structure
of state (\ref{eq:ALsat_state}) is not robust against small correction
to the exact saturation \citep{HaydenJozsaPetzWinter04:StructureofStates}.
The saturation of the EWCS or the EWMI should also be found only in
the large $N$ limit. We expect, however, that our conclusion itself
still survives: the EWCS and the EWMI with some appropriate quantum
corrections will still capture classical correlations. 

The bit thread formalism \citep{FreedmanHeadrick17:BitThreads} has
been cooperative with these holographic optimized correlation measures.
The bit threads for the bipartite EWCS was discussed in \citep{AgonDeBouerPedraza18:GeometricAspectsofHolographicBitThreads,DuChenShu19:BitthreadsandholographicEntanglementOfPurification:2019emy,BaoChatwinPollack19:TowardsaBitThreads,HarperHeadrick19:BitthrreadsandHolographicEntanglementofPurification}
and generalized to the multipartite EWCS \citep{HarperHeadrick19:BitthrreadsandHolographicEntanglementofPurification}.
It will be interesting to seek a bit thread formalism for $E_{M}$
as well. Also, a multipartite generalization of $E_{M}$ would provide
us a new tool to probe a specific aspect of the holographic correlations. 

\subsection*{Acknowledgments}

We are grateful to Arpan Bhattacharyya, Souvik Dutta, Jonathan Harper,
Masamichi Miyaji, Yoshifumi Nakata, Tadashi Takayanagi, and Yang Zhou
for valuable comments on a draft of the present paper. We thank Jesse
C. Cresswell and Massimiliano Rota for a lot of stimulating discussions
which essentially motivated this work. We are grateful to \textquotedbl Quantum
Information and String Theory 2019\textquotedbl{} at the Yukawa Institute
for Theoretical Physics in Kyoto University where this work was initiated.
We also thank ``Gravity - New perspectives from strings and higher
dimensions 2019'' at the Centro de Ciencias de Benasque Pedro Pascual,
where this work was completed. K.U. is supported by Grant-in-Aid for
JSPS Fellows No.18J22888.

\section{Appendix}

\subsection{The Multipartite Generalization}

In this appendix, we complement some missing pieces in the previous
study \citep{UmemotoZhou18:Entanglement} of the multipartite generalization
of the mutual information, the EOP, and the squashed entanglement
as well as their holographic duals. The mutual information $I(A:B)$
has various multipartite generalizations. One of them is called the
total correlation defined by 
\begin{align}
T_{n}(A_{1}:\cdots:A_{n}) & :=S(\rho_{A}||\rho_{A_{1}}\otimes\cdots\otimes\rho_{A_{n}})\\
 & =\sum_{i=1}^{n}S_{A_{i}}-S_{A}\\
 & =I(A_{1}:A_{2})+I(A_{1}A_{2}:A_{3})\nonumber \\
 & \ +\cdots+I(A_{1}\cdots A_{n-1}:A_{n}),
\end{align}
where $S(\rho||\sigma)={\rm Tr}\rho(\log\rho-\log\sigma)$ is the
relative entropy. There is another generalization called the dual
total correlation 
\begin{align}
D_{n}(A_{1}:\cdots:A_{n}) & :=S_{A_{1}\cdots A_{n}}-\sum_{i=1}^{n}S(A_{i}|A_{1}\overset{i}{\check{\cdots}}A_{n})\\
 & =I(A_{1}:A_{2}\cdots A_{n})+I(A_{2}:A_{3}\cdots A_{n}|A_{1})\nonumber \\
 & \ +\cdots+I(A_{n-1}:A_{n}|A_{1}\cdots A_{n-2}),
\end{align}
where $S(A|B)=S_{AB}-S_{B}$ is the conditional entropy and $\overset{i}{\check{\cdots}}$
denotes the exclusion of $A_{i}$. The $T_{n}$ and $D_{n}$ are monotonically
non-increasing under strict local operations, vanish if and only if
the state is totally decoupled, and $T_{n}=D_{n}=\sum_{i=1}^{n}S_{A_{i}}$
if the state is pure. 

A multipartite generalization of the EOP \citep{UmemotoZhou18:Entanglement,BaoHalpern18:Conditional}
and the squashed entanglement \citep{YangHorodeckiHorodeckiHorodeckiOppenheimSong09:Squashed,AvisHaydenSavov08:DistributedCompressionandMultipartySquashedEntanglement}
are given as follows:

\begin{align}
E_{P}(A_{1} & :\cdots:A_{n})=\frac{1}{2}\min_{\ket{\psi}_{A_{1}A'_{1}\cdots A_{n}A'_{n}}}T_{n}(A_{1}A'_{1}:\cdots:A_{n}A'_{n}).\label{eq:MEOP}\\
E_{sq}(A_{1} & :\cdots:A_{n})=\frac{1}{2}\min_{\rho_{A_{1}\cdots A_{n}E}}T_{n}(A_{1}:\cdots:A_{n}|E),
\end{align}
where $T_{n}(A_{1}:\cdots:A_{n}|E)=I(A_{1}:A_{2}|E)+I(A_{1}A_{2}:A_{3}|E)+\cdots+I(A_{1}\cdots A_{n-1}:A_{n}|E)$.
The multipartite EOP is monotonically non-increasing under strict
local operations. The holographic dual of the multipartite EOP was
proposed as the multipartite EWCS \citep{UmemotoZhou18:Entanglement}. 

We can generalize the discussion of the holographic dual of the bipartite
squashed entanglement as follows. The multipartite squashed entanglement
can be written as 
\begin{align}
E_{sq} & =\frac{1}{2}T_{n}(A_{1}:\cdots:A_{n})+\frac{1}{2}\min_{\rho_{A_{1}\cdots A_{n}E}}Q_{n}(A:E),
\end{align}
where we define $Q_{n}(A;E)\coloneqq I(A_{1}\cdots A_{n}:E)-\sum_{i=1}^{n}I(A_{i}:E).$
This can be both positive and negative in generic quantum system.
In holography, however, the monogamy of mutual information implies
$Q_{n}\geq0$. For $n=2$, it reproduces the non-positivity of tripartite
information $Q_{2}(AB;E)=I(AB:E)-I(A:E)-I(B:E)=-I_{3}(A:B:E)\geq0$.
It again results in a conjecture that holographic multipartite squashed
entanglement is equivalent to half of the total correlation, 
\begin{equation}
E_{sq}(A_{1}:\cdots:A_{n})=\frac{1}{2}T_{n}(A_{1}:\cdots:A_{n}).
\end{equation}

For the latter convenience, we introduce two non-negative quantities
for $n\geq3$: 
\begin{equation}
X_{n}:=\frac{(n-1)T_{n}-D_{n}}{n-2},\ Y_{n}:=\frac{(n-1)D_{n}-T_{n}}{n-2}.
\end{equation}
They are normalized so that $X_{n}=Y_{n}=\sum_{i=1}^{n}S_{A_{i}}$
holds for pure states. They are positive semi-definite as it is clear
from the following expressions,

\begin{align}
X_{n}(A_{1}:\cdots:A_{n}) & =\frac{1}{n-2}\sum_{i=1}^{n}T_{n-1}(A_{1}:\overset{i}{\check{\cdots}}:A_{n}),\\
Y_{n}(A_{1}:\cdots:A_{n}) & =\frac{1}{n-2}\sum_{i=1}^{n}D_{n-1}(A_{1}:\overset{i}{\check{\cdots}}:A_{n}|A_{i}),
\end{align}
where $D_{n}(A_{1}:\cdots:A_{n}|E)=I(A_{1}:A_{2}\cdots A_{n}|E)+I(A_{2}:A_{3}\cdots A_{n}|A_{1}E)+\cdots+I(A_{n-1}:A_{n}|A_{1}\cdots A_{n-2}E)$.
$X_{n}$ is monotonically non-increasing under strict local operations,
while the $Y_{n}$ is not necessarily. Both $X_{n}$ and $Y_{n}$
are not faithful i.e. there exists a state which is not decoupled
$\rho_{A}\neq\rho_{A_{1}}\otimes\cdots\otimes\rho_{A_{n}}$ but $X_{n}=0$
or $Y_{n}=0$. Thus we do not consider each of them as a good correlation
measure. Note a balance equation,
\begin{align}
T_{n}+D_{n} & =X_{n}+Y_{n}=\sum_{i=1}^{n}I(A_{1}:\overset{i}{\check{\cdots}}:A_{n}).
\end{align}

For holographic states, the monogamy of mutual information leads to
a generic ordering 
\begin{equation}
X_{n}\leq T_{n}\leq D_{n}\leq Y_{n}.\label{eq:XTDY}
\end{equation}
Indeed, $T_{n}\leq D_{n}$ follows from the monogamy of mutual information,
\begin{align}
 & D(A_{1}:\cdots:A_{n})\nonumber \\
 & =I(A_{1}:A_{2}\cdots A_{n})+I(A_{2}:A_{3}\cdots A_{n}|A_{1})\nonumber \\
 & \ \ +\cdots+I(A_{n-1}:A_{n}|A_{1}\cdots A_{n-2})\nonumber \\
 & \geq I(A_{1}:A_{2}\cdots A_{n})+I(A_{2}:A_{3}\cdots A_{n})\nonumber \\
 & \ \ +\cdots+I(A_{n-1}:A_{n})\nonumber \\
 & =T(A_{1}:\cdots:A_{n}).
\end{align}
Then $X_{n}\leq T_{n}$ and $D_{n}\leq Y_{n}$ are obvious by their
definition.

Now we present some lower bounds on the multipartite EOP, which generalizes
and complements the inequalities proven in \citep{UmemotoZhou18:Entanglement}.
The multipartite EOP is bounded from below by half of any multipartite
correlation measure $\Theta$ which satisfies (i) $\Theta=\sum_{i=1}^{n}S_{A_{i}}$
for pure $n$-partite states, and (ii) is non-increasing under strict
local operations, 
\begin{equation}
E_{P}\geq\frac{1}{2}\Theta.
\end{equation}

It is obvious from the definition (\ref{eq:MEOP}) following the same
logic as in \citep{UmemotoZhou18:Entanglement}. Here $T_{n}$, $D_{n}$,
and $X_{n}$ satisfy both conditions, but $Y_{n}$ does not satisfy
(ii). Thus, we get three inequalities for generic multipartite states
\begin{equation}
E_{P}\geq\frac{1}{2}\max\{X_{n},T_{n},D_{n}\}.
\end{equation}
The lower bounds by $T_{n}$ and $X_{3}$ were proven in \citep{UmemotoZhou18:Entanglement},
and the above inequality gives $n$-partite generalization for $X_{n}$.
On the other hand, the bound by $D_{n}$ is totally new. Interestingly,
$D_{n}$ gives a stricter lower bound on $E_{P}$ than $T_{n}$ in
holography by the ordering (\ref{eq:XTDY}). One can check that $E_{W}\geq\frac{1}{2}D_{n}$
always holds, while $E_{W}\geq\frac{1}{2}Y_{n}$ is not true in general. 

\bibliographystyle{unsrt}
\bibliography{all4th}

\end{document}